\documentclass[fleqn,usenatbib]{mnras}

\usepackage{newtxtext,newtxmath}

\usepackage[T1]{fontenc}

\DeclareRobustCommand{\VAN}[3]{#2}
\let\VANthebibliography\thebibliography
\def\thebibliography{\DeclareRobustCommand{\VAN}[3]{##3}\VANthebibliography}

\usepackage{graphicx}	
\usepackage{amsmath}	

\usepackage{xcolor}
\usepackage{subcaption}

\title[Mixing-induced cooling]{Mixing-induced thermal instabilities and coronal condensations}

\author[B. Snow \& A. Hillier]{
B. Snow,$^{1}$\thanks{E-mail: b.snow@exeter.ac.uk}
and A. Hillier$^{1}$
\\
$^{1}$University of Exeter, UK
}

\date{Accepted XXX. Received YYY; in original form ZZZ}

\pubyear{2021}

\begin{document}
\label{firstpage}
\pagerange{\pageref{firstpage}--\pageref{lastpage}}
\maketitle

\begin{abstract}
     {Cool, dense material is frequently observed to permeate the hot, tenuous solar corona in the form of prominences, spicules and coronal rain. Both the cool material and surrounding corona exist at temperatures that are effectively thermally stable, in that their local radiative losses occur on relatively long timescales compared to the dynamics. However, the intermediate temperatures at the boundaries between cool and hot material are subject to highly efficient radiative losses. The turbulent motions in the solar atmosphere can drive mixing between condensations and the corona, leading to the formation of intermediate temperatures and thus efficient cooling.
   Here, a 3D radiative MHD simulation is performed of the shear-driven Kelvin-Helmholtz Instability (KHI) occurring between a cool condensation and the hot solar corona. 
   During the evolution, thermal instabilities form naturally within the mixing layer, and grow with time to produce long, narrow structures that extend perpendicular to the magnetic field. The thermal instabilities form self-consistently within the mixing layer as small isolated events, and are then stretched by the background flows to create long structures in relatively narrow planes. The turbulent flows agitate the condensations and cause them to fragment, creating smaller localised clumps of cool, dense (prominence-like) material that can merge and act as seeds for further condensations. The thermal instabilities act to replenish the cool, dense material lost due to mixing, with the total mass of cool material being approximately constant through time and are found to account for 15-20\% of all radiative losses in the turbulent plasma. As such, thermal instabilities are dynamically important features of condensation-corona mixing, can self-consistently arise from radiative losses and turbulent motions, and can create additional cool, dense material.
   }
   {}
   {}
   {   }
\end{abstract}



\section{Introduction}


The solar corona is a hot ($T\approx10^6$K), tenuous medium that is permeated by cool ($\approx10^4$K), dense, chromospheric-like material in the form of spicules \citep{Beckers1968,Sterling2000,Periera2014}, prominences/filaments \citep{1995ASSL..199.....T,2010SSRv..151..333M,Parenti2014} and coronal rain \citep{2001SoPh..198..325S,Hinode2019,Antolin2020}. The density within these cool structures can be several orders of magnitude higher than the surrounding coronal material, in addition to the temperature being several orders of magnitude lower. At the boundary between the cool and hot material, a transition region of intermediate temperature and density forms. For prominences, this is known as the prominence-corona transition region (PCTR) \citep{1976SoPh...50..365O,2004SoPh..223...95C,Parenti2014}.

The interface between condensations and the solar corona can become unstable to mixing instabilities that are excited by the intrinsic turbulent motions of the solar atmosphere, with the Kelvin Helmholtz Instability (KHI) being a key instability that develops in this region \citep[e.g.][]{Antolin2015ApJ...809...72A, Antolin2018ApJ...856...44A}. Initial interest in this development of the KHI was in the context of a heating mechanism due to the dissipation of kinetic and magnetic energy through the turbulence this instability creates \citep{Antolin2015ApJ...809...72A}. However, recent studies have shown that the turbulence develops a layer where the two distinct phases are being mixed to dynamically develop a PCTR, and in this layer cooling process due to radiative losses can play a dominant role in the thermal evolution of the system. \citep{Hillier2023}.

Both the corona ($\approx10^6$K) and the cool material ($\approx 10^4$K) are relatively stable to radiative losses, in that the timescale of losses is significantly longer than the dynamic timescales. However, the intermediate temperatures ($\approx 10^5$K) that occur in the mixing layer are susceptible to efficient radiative losses that can extract significantly more thermal energy than is deposited through turbulent heating on timescales that are shorter than the timescales of the dynamics \citep{Hillier2019, Hillier2023}. Thus, the radiative losses of the mixing layer can induce cooling in condensation-corona systems. The observational signatures of cooling may manifest as an increased emission in transition-region lines \citep{Snow2025}, which is often interpreted as a sign of heating. The mixing induced-cooling process discussed here is understood to occur in many different astrophysical systems \citep[e.g.][]{1990MNRAS.244P..26B,2020ApJ...894L..24F}.

Optically-thin radiative losses are efficient at the cooling mixing layers around prominences, coronal rain and spicules and are known to preclude the existence of thermal instabilities \citep{Hermans2021}. The thermal instability is thought to be a key driving mechanism behind the cool, dense condensations that form in the solar atmosphere \citep{Field1965,Claes2019}, intergalactic medium \citep{Jennings2021MNRAS.505.5238J}, and molecular outflows of galaxies \citep{Gronke2018}. Thermal instabilities in low plasma-$\beta$ MHD are seen to evolve as long, thin structures perpendicular to the magnetic field \citep{Claes2020}.
The physical process behind this instability is that the material locally cools due to radiative losses, and the rate of thermal energy loss is faster than the rate at which thermal conduction can transport heat into the region. When this occurs, the instability is generally categorised as being isobaric (constant pressure) or isochoric (constant density) cooling \citep{Field1965}. Isochoric cooling requires cooling to occur so fast that sound waves have insufficient time to travel across the system to give causal contact between the condensation and the surroundings, which is not expected for condensations in the solar atmosphere implying approximately isobaric cooling is expected \citep{Field1965,Claes2019}. For isobaric cooling, the losses remove thermal energy, and hence pressure, from the system. As such, a low pressure region forms which leads to pressure-gradient driven flows that bring material into the cooled region. This increases the pressure and hence temperature leading to the material being subject to further cooling. This positive feedback cycle may lead to cool, dense regions forming in a mixing layer.

Despite thermal instabilities being a fundamental feature of radiative MHD, they were not seen to form in 2D KHI mixing simulations of the corona-condensation interface, e.g., \citet{Hillier2023}, where  
cool material formed in the mixing layer due to radiative losses. 
The cool material they found was not dense, implying isochoric cooling. 
However, no thermal instability-like positive-feedback mechanism was present, and the formation of cool material was purely a consequence of thermal non-equilibrium. 
As such, the mixing did not produce any prominence-like material which is both cool and dense. It is likely that the out-of-plane magnetic field in their set-up prevented the onset of thermal instabilities since compressible flows along the magnetic field were not possible. Since thermal instabilities naturally arise and are critical in creating cool, dense mass in other astrophysical systems, e.g. the  cloud-crushing events modelled by \citet{Gronke2018}, it stands to reason that thermal instabilities may also generate cool, dense material in mixing layers between the cool and hot phases of the solar corona, but this will be a fully 3D effect.  





Here, a 3D radiative MHD simulation is performed of the Kelvin-Helmholtz Instability to investigate the formation of thermal instabilities in condensation-corona mixing events. No initial thermal instability is prescribed and the system is left to evolve freely. The results indicate that the thermal instability can naturally arise in mixing scenarios to produce cool, dense material, similar to coronal condensations. 

\section{Methodology}

Numerical simulations are performed using the single-fluid MHD mode of the (P\underline{I}P) code \citep{Hillier2016} that solves the MHD equations in conservative form. The fourth-order, central difference solver of (P\underline{I}P) is used. The specific set of non-dimensional equations we use are given by:
\begin{gather}
\frac{\partial \rho }{\partial t} + \nabla \cdot (\rho \textbf{v}) = 0 \label{eqn:plasma1}\\
\frac{\partial}{\partial t} (\rho \textbf{v})+ \nabla \cdot \left( \rho \textbf{v} \textbf{v} + P \textbf{I} - \textbf{B B} + \frac{\textbf{B}^2}{2} \textbf{I} \right) =0\\
\frac{\partial}{\partial t} \left( e + \frac{\textbf{B}^2}{2} \right) + \nabla \cdot \left[ \textbf{v} ( e + P) -  (\textbf{v} \times \textbf{B}) \times \textbf{B} \right] =-\rho ^2 \Lambda(T), \label{eqn:energy}\\
\frac{\partial \textbf{B}}{\partial t} - \nabla \times (\textbf{v} \times \textbf{B}) = 0, \\
e = \frac{P}{\gamma -1} + \frac{1}{2} \rho  v ^2, \\
\nabla \cdot \textbf{B} = 0,\label{eqn:plasma2}
\end{gather}
for density $\rho$, energy $e$, pressure $P$, velocity $\textbf{v}$ and $\textbf{B}$ is the magnetic field divided by $\sqrt{4 \pi}$. The adiabatic index $\gamma=5/3$ and is constant. Temperature and pressure are related using the non-dimensional ideal gas law:
\begin{gather}
    T=10^6\frac{\gamma P}{\rho}.
\end{gather}
Note that $T$ multiplied by unit Kelvin gives the dimensionalised temperature.

\subsection{Radiative losses}

\begin{figure}
    \centering
    \includegraphics[width=0.95\linewidth]{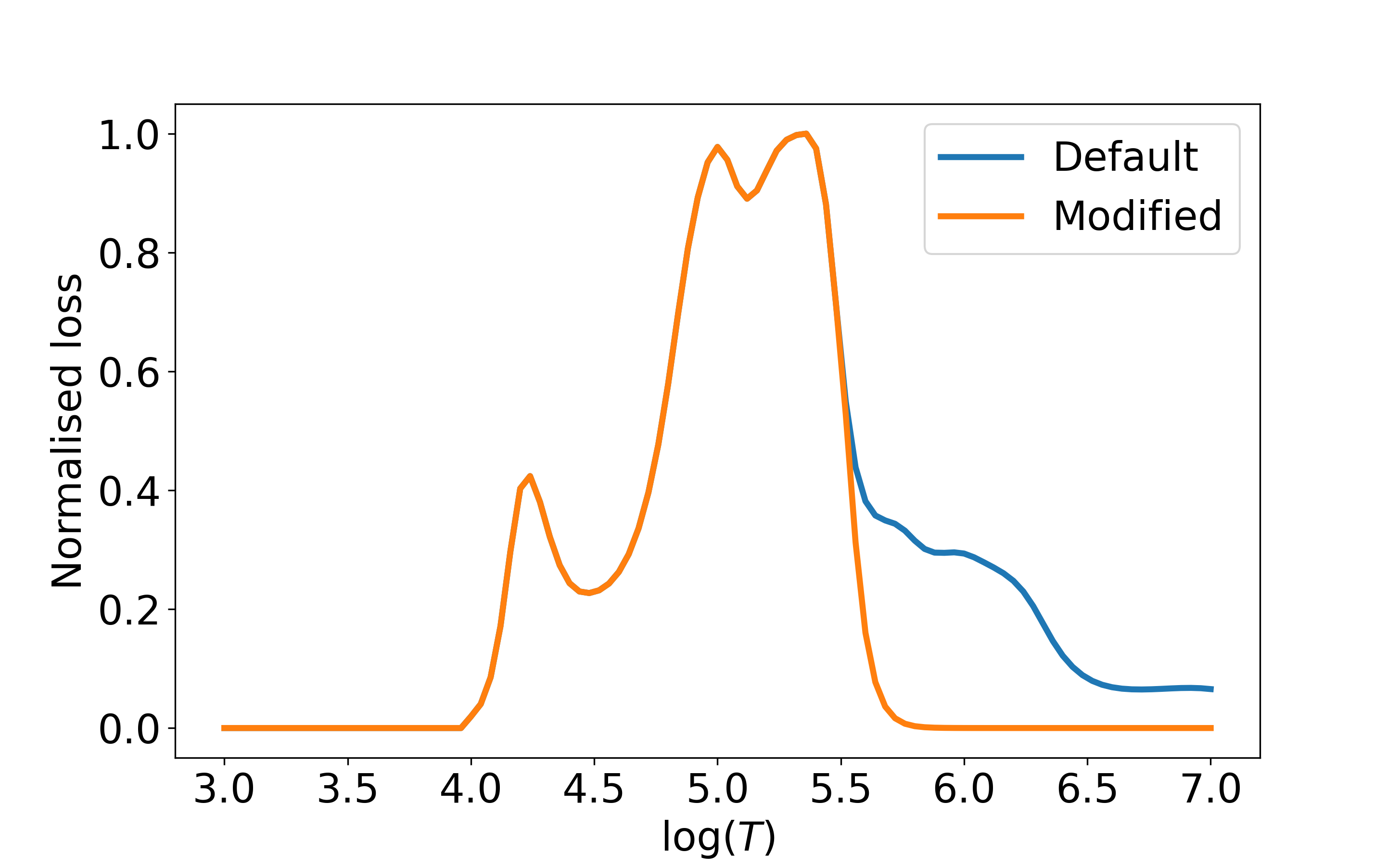}
    \caption{Normalised loss function $\Lambda (T)$. The blue line shows the default Chianti generated loss curve with the default abundance file. The orange line shows the modified loss curve used in this work which has zero losses above $10^6$K. Note that the maximum cooling rate in this paper is set to $\max (\Lambda (T))=10^{-3}$}.
    \label{fig:lossfunc}
\end{figure}

A cooling term of the form $\rho^2 \Lambda(T)$ is included to model the radiative losses in the solar atmosphere, which acts as an energy sink in Equation \ref{eqn:energy}. The temperature-dependent cooling profile is generated using Chianti v10 \citep{Dere1997,Zanna2021} to calculate the radiative losses as a function of temperature, as shown in Figure \ref{fig:lossfunc}.  

The Chianti loss curve is modified such that the losses are effectively zero above $T=10^6$ according to 
\begin{gather}
    \Lambda (T)= 
\begin{cases}
    0,& \text{if } T \leq {9575}\\
    10^{-3}\times\Lambda_0 (T),& \text{if }  {9575} <T < 10^{5.5}\\
    10^{-3}\times\Lambda_0 (T)(1-\tanh\left( \left( \frac{\log_{10}(T)-5.5}{0.4}\right)^2 \right),  & \text{otherwise}
\end{cases}
\end{gather}
where $\Lambda_0(T)$ is the normalised temperature-dependent cooling function from Chianti. The modification assumes no radiative losses in the upper layer. The implicit assumption here is that some unmodelled background process maintains the temperature of the solar corona. {Similarly, radiative cooling is identically zero below 9575.}


\subsection{Initial conditions}

\begin{table}
    \centering
    \caption{Initial conditions in the lower (coronal) and upper (prominence) layers}
    \begin{tabular}{c|c|c}
        Variable & lower & upper \\ \hline
        $\rho$ & 1 & 100 \\
        $p$ & $1/\gamma$ & $1/\gamma$ \\
        $v_x$ & $\approx 0.198$ & $\approx-0.00198$ \\
        $v_y,v_z$ & 0 & 0 \\
        $B_x,B_y$ & 0 & 0 \\
        $B_z$ & $\approx 4.90$ & $\approx 4.90$ \\
        $T$ & $10^6$ & $10^4$ \\
        $c_s$ & 1 & 0.1 \\
        plasma-$\beta$ & 0.05 & 0.05 
    \end{tabular}
    \label{tab:initialconditions}
\end{table}

We model the Kelvin-Helmholtz Instability (KHI) which is driven by a shear layer at the interface between two regions of different density. The initial conditions are given in Table \ref{tab:initialconditions}. The upper layer mimics a dense prominence-like region, whereas the lower layer is coronal-like. The density changes by two orders of magnitude across the interface with the pressure constant. As such, the temperature also changes by a factor of 100 between the two layers. The magnetic field is initially in the $z-$direction only with $B_z \approx 4.90$, with $B_x=B_y=0$ initially, corresponding to an initial plasma-$\beta=\frac{P}{B^2/2}=0.05$, i.e., the ratio of gas pressure to magnetic pressure. 

Normalisation of the temperature is chosen such that a simulation value of $T=10^6$ corresponds to a temperature of $10^6$K, i.e., the sparse layer is coronal-like temperatures ($10^6$K), and the dense layer is prominence-like ($10^4$K). Radiative losses are modelled using the $\rho^2 \Lambda (T)$ term in the energy equation and the timescale for radiative cooling is scaled such that for a density of unity, the timescale for energy loss is $10^3$ time units at the peak of the radiative loss curve. As the mixing evolves, the density squared dependence of the cooling rate leads to far more efficient radiative losses. 

The KHI is triggered by applying a shear flow of the form 
\begin{gather}
    v_x = \begin{cases}
        0.2\frac{\rho_u}{(\rho_u+\rho_l)} \approx 0.198 & \text{if } y<0 \\
        -0.2\frac{\rho_l}{(\rho_u+\rho_l)} \approx -0.00198 & \text{if } y>0
    \end{cases}
\end{gather}
where $\rho_l=1,\rho_u=100$ are the densities of the lower and upper layers respectively. The sheer is located at the interface between the corona and prominence material. To assist in the formation of turbulence, a small random perturbation is applied to the $v_y$ velocity with a magnitude of $\pm0.005$ throughout the full domain. The $z-$component of velocity is set to $v_z=0$ initially.

The domain size spans $-1.5\leq x\leq 1.5, -2.5\leq y<0.5, -75 \leq z \leq 75$. The $z-$ direction needs to be significantly larger than the $x,y-$directions due to the propagation of Alfv\'en waves along the magnetic field (which is originally in the $z-$direction only). The grid is resolved by $600\times600\times750$ cells in the $x,y,z-$ direction respectively, with $\Delta x= \Delta y = 0.005, \Delta z =0.2$. This is a relatively large aspect {ratio}, however the high Alfv\'en speed in the $z-$direction means that there is approximate coherency of Alfv\'en waves at the grid scale, as discussed in Section \ref{Sec:aspect}. The boundary conditions are set to periodic in $x,z$. The boundary conditions in $y$ {are set to mirror boundaries} where values are reflected across the interface (e.g., $\rho \left[ -1\right]=\rho \left[ 1\right]$), and the velocity and magnetic fields through the boundary are reflected (e.g., $v_y \left[ -1\right]=-v_y \left[ 1\right]$). The simulation is stopped before any mixing has occurred near the $y$ boundaries to prevent the influence of the boundaries. The simulations are performed using 2800 CPUs on the COSMA7 machine (part of the DiRAC Consortium), with the domain decomposed in $8\times 14 \times 25$ (x,y,z) processor blocks.

\subsection{Physical interpretation of this setup}

\begin{table}
    \centering
    \caption{Normalisation constants}
    \begin{tabular}{c|c}
        Variable & value \\ \hline
        $\hat{T}$ & 1K \\
        $\hat{c}_s$ & 166 km/s \\
        $\hat{L}$ & 100 km \\
        $\hat{t}$ & 0.6 s 
    \end{tabular}
    \label{tab:normalisationconstants}
\end{table}

To consider the simulation in the context of the solar atmosphere, the dimensional length and time scales of the system can be calculated. For the solar corona, a typical cooling time is on the order of 2000$\,$s. From the original cooling curve in Figure \ref{fig:lossfunc} (blue line), the loss rate at $T=10^6$ is approximately $\Lambda(T)=0.3 \hat{\Lambda}$, where $\hat{\Lambda}=10^{-3}$ is the peak cooling rate in our simulation. Therefore, our dimensionless cooling rate is approximately $3\times 10^{-4}$. Now, multiplying a typical cooling time \citep[2000 s, ][]{Cargill2015RSPTA.37340260C} by the cooling rate gives the dimensional simulation time step as $\hat{t}\approx 0.6\,$s. 
The coronal sound speed can be estimated as $\hat{c}_s\approx166\sqrt{\hat{T}}\,\mbox{m/s}=166\,$km/s \citep{Priest1982}. The length scale is then $\hat{c}_s\times \hat{t}\approx100\,$km, meaning that our simulation is approximately a $300\times 300 \times 15000\,$km box. Changing the cooling rate would effectively change the length scale of the system, i.e., scaling the loss curve to $10^{-2}$ would result in a box an order of magnitude larger in each direction.  A table summarising the normalisation constants is given in Table \ref{tab:normalisationconstants}.



\section{Results}

\subsection{Simulation evolution}

\begin{figure}
    \centering
    \includegraphics[width=0.99\linewidth]{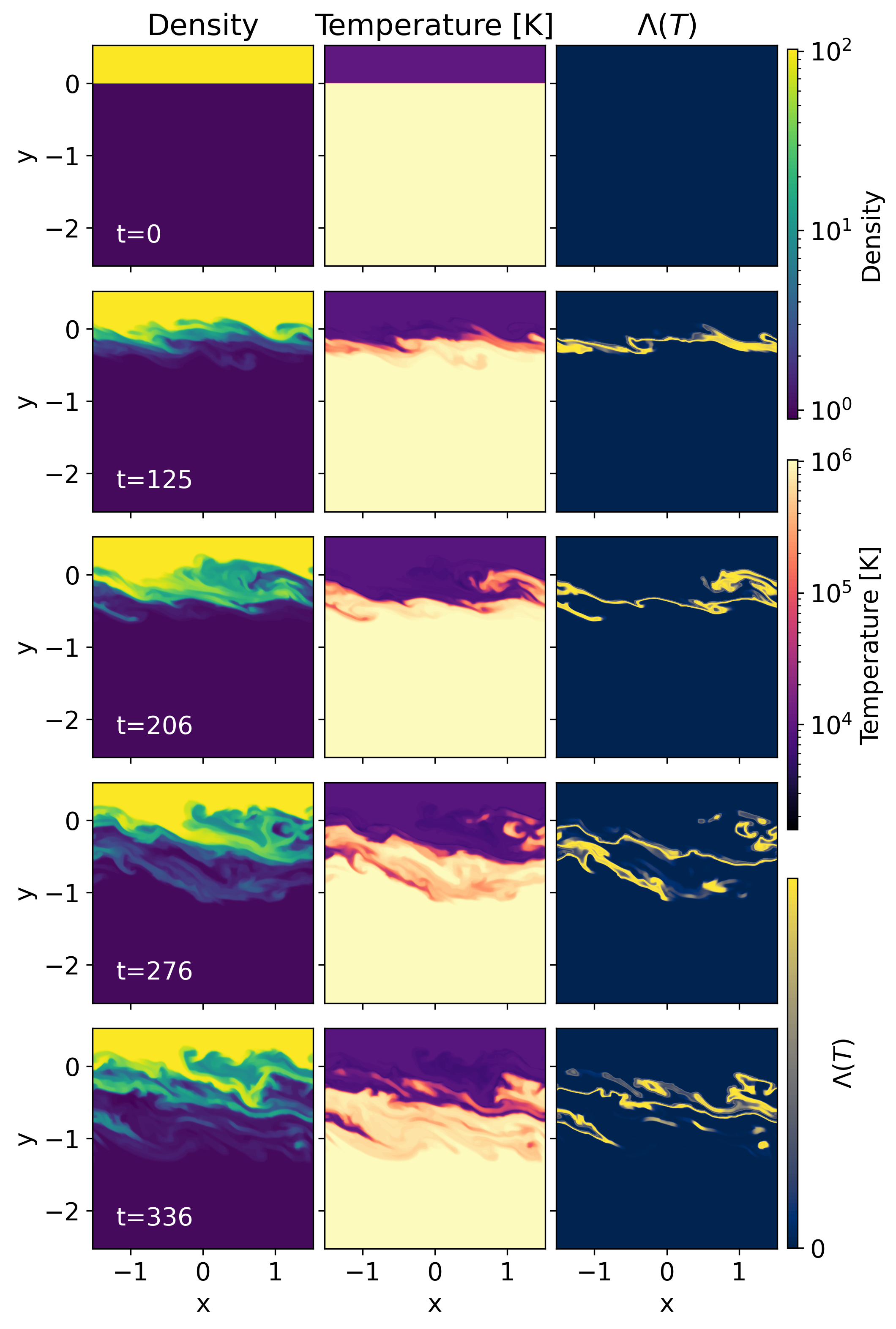}
    \caption{Slice of the simulation at $z=0$ at different times showing the density (left), temperature (centre) and losses (right).}
    \label{fig:simevo}
\end{figure}

\begin{figure*}
    \centering
    {\includegraphics[width=0.75\linewidth]{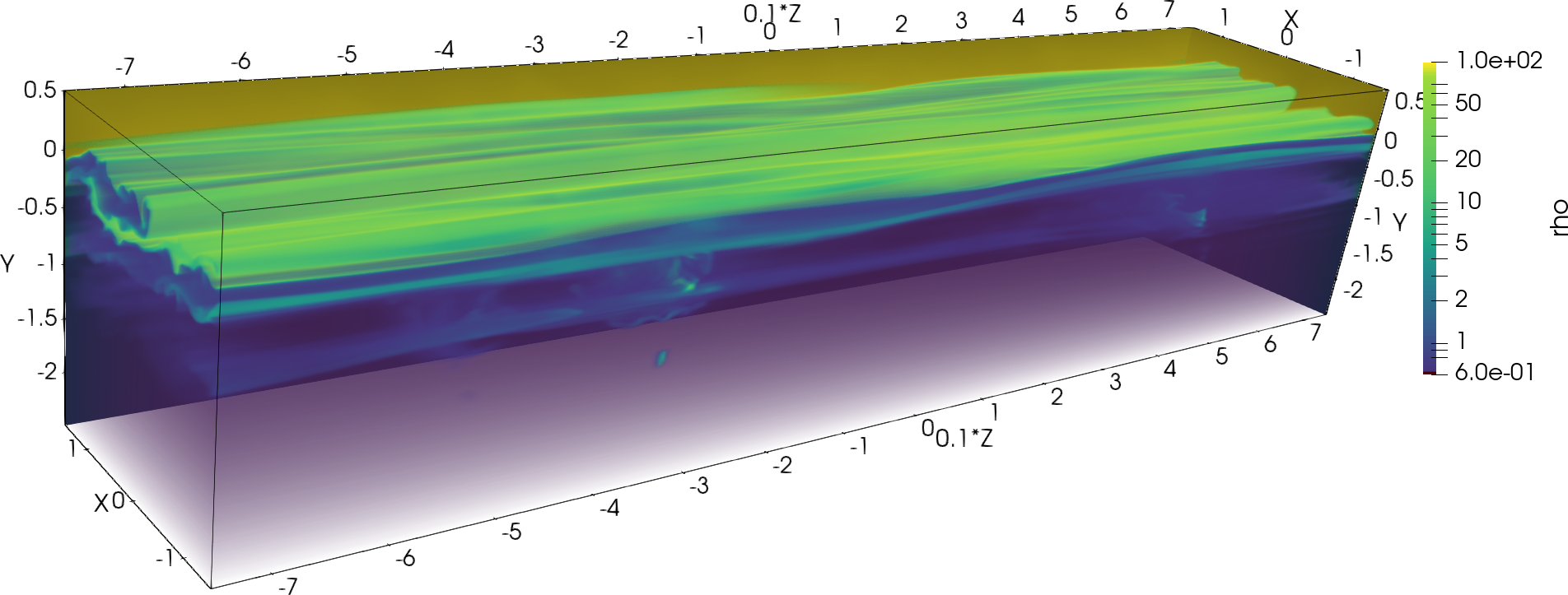}\label{fig:volume:sub1}}\\
    {\includegraphics[width=0.75\linewidth]{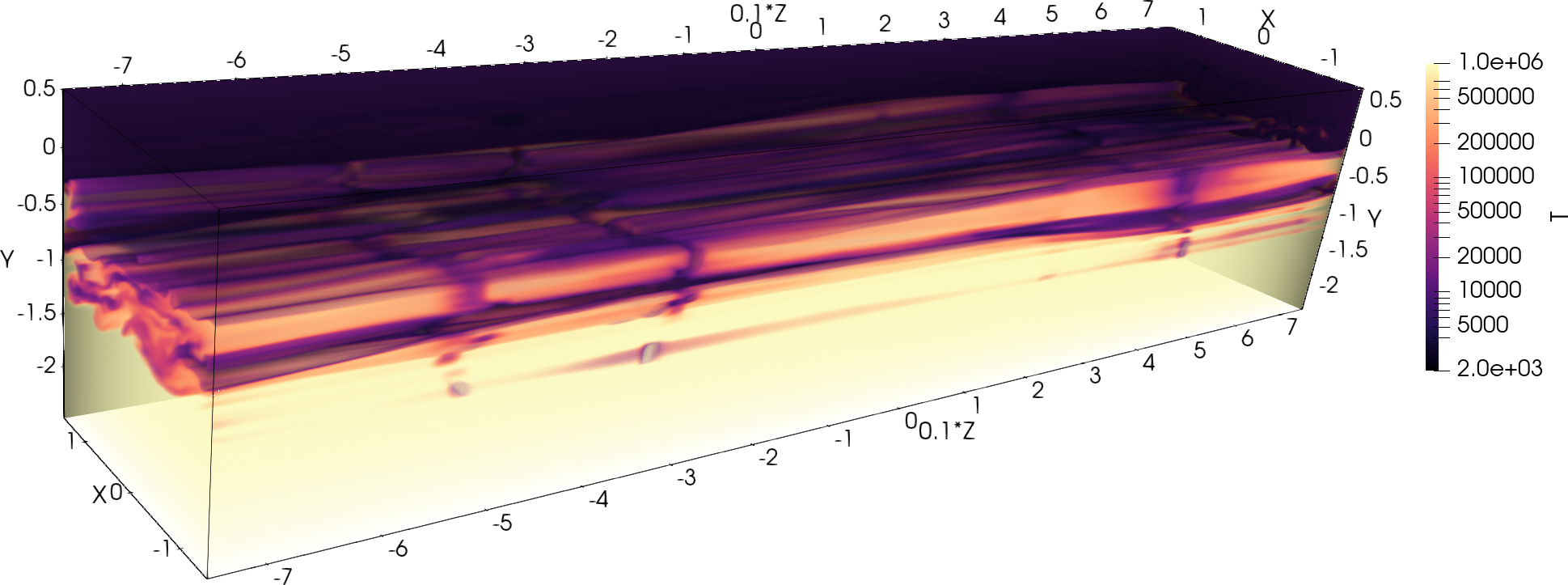}\label{fig:volume:sub2}}\\
    {\includegraphics[width=0.75\linewidth]{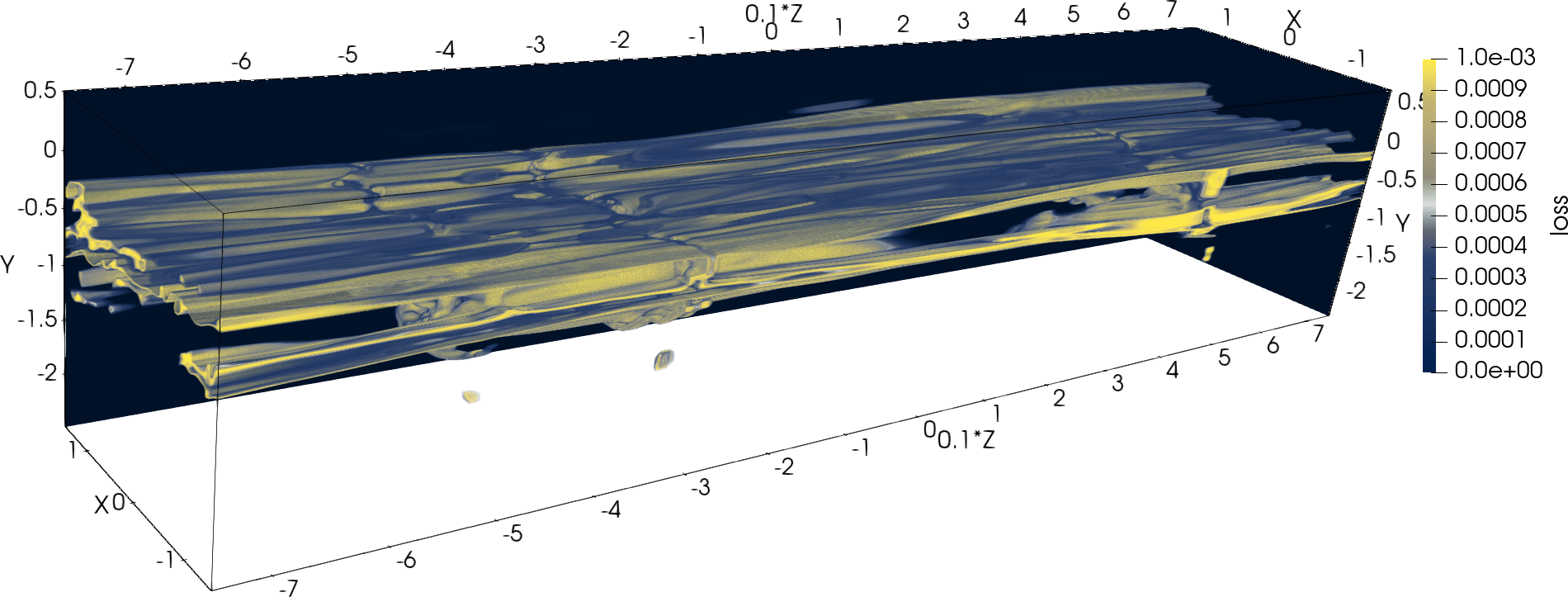}\label{fig:volume:sub3}}\\
    \includegraphics[width=0.75\linewidth]{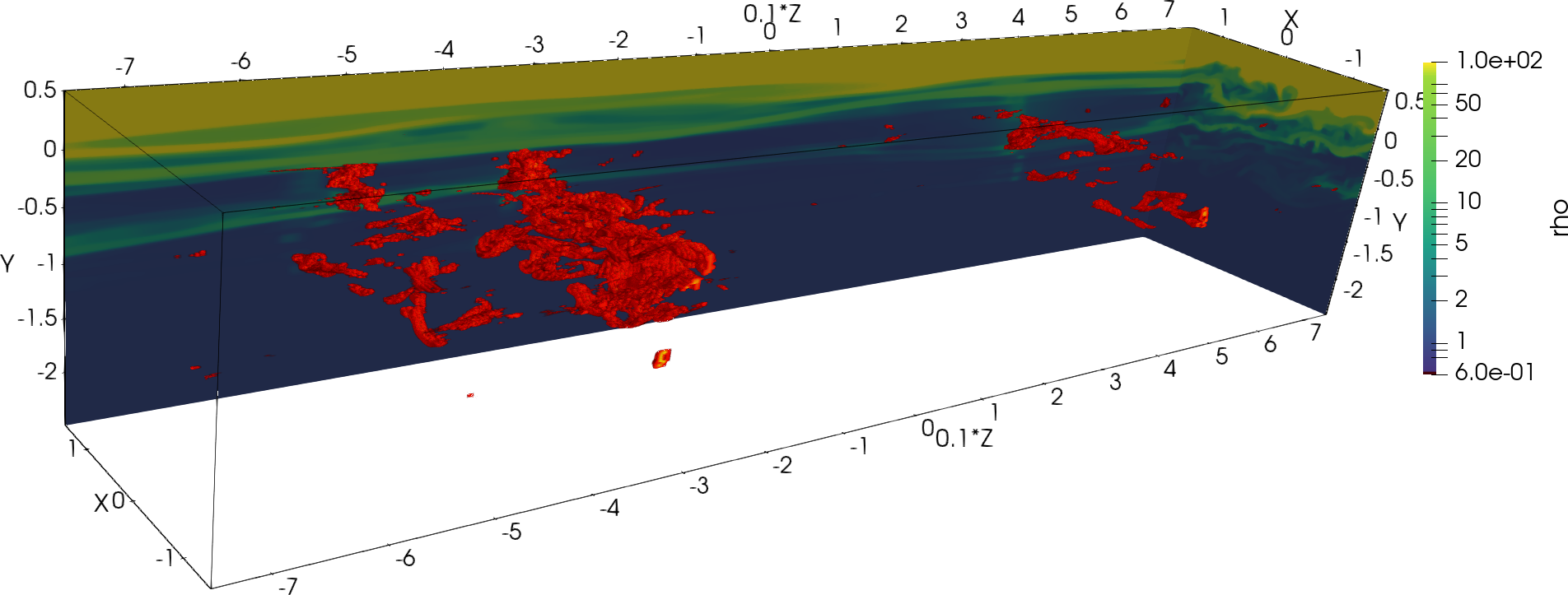}
    \caption{3D volume plots at time $t=336$ of density (a), temperature (b), losses (c). Panel (d) shows the identified thermal instability-like structures with arbitrary colouring, with the density as the background colour. Note that the opacity of the values outside of the mixing layer have been set to zero.}
    \label{fig:volume3D}
\end{figure*}

The shear flow across the interface lead to the formation of the Kelvin-Helmholtz instability, mixing the prominence and coronal material. A 2D slice of the domain at $z=0$ is shown in Figure \ref{fig:simevo}, with 3D volume renderings of the mixing layer shown in Figure \ref{fig:volume3D}. The mixed material exists at temperatures that are subject to efficient radiative losses, leading to cooling occurring within the mixing layer. The cooling results in a mixing layer that has diffuse density, and sharper temperature gradients, as seen in Figure \ref{fig:simevo} and previous works in 2D \citep{Hillier2023,Snow2025}. Radiative losses are centred in the mixing layer, as expected.  

The 3D renderings in Figure \ref{fig:volume3D} show the density (Figure \ref{fig:volume3D}a), temperature (Figure \ref{fig:volume3D}b) and temperature-dependent loss component ($\Lambda(T)$, Figure \ref{fig:volume3D}c). 
The 3D renderings show the turbulent nature of the mixing in all 3 directions, and thus the mixing is fundamentally different to the 2D models. A detailed discussion of the 3D-ness of the turbulence is included in Section \ref{sec:3Dness} Figures \ref{fig:volume3D}a and b show that the density and temperature of the mixing layer is broadly distributed, with the turbulent mixing layer spanning the full extent of the $x,z$ directions. Similarly, the losses (Figure \ref{fig:volume3D}c) are turbulent and reasonably well-distributed through the mixing layer. An aspect that is present in these 3D simulations is thermal instabilities (shown in Figure \ref{fig:volume3D}d), which are not possible to form in the 2D simulations.



\subsection{Thermal instabilities} \label{sec:thermal_instabilities}


\begin{figure}
    \centering
    \includegraphics[width=0.95\linewidth]{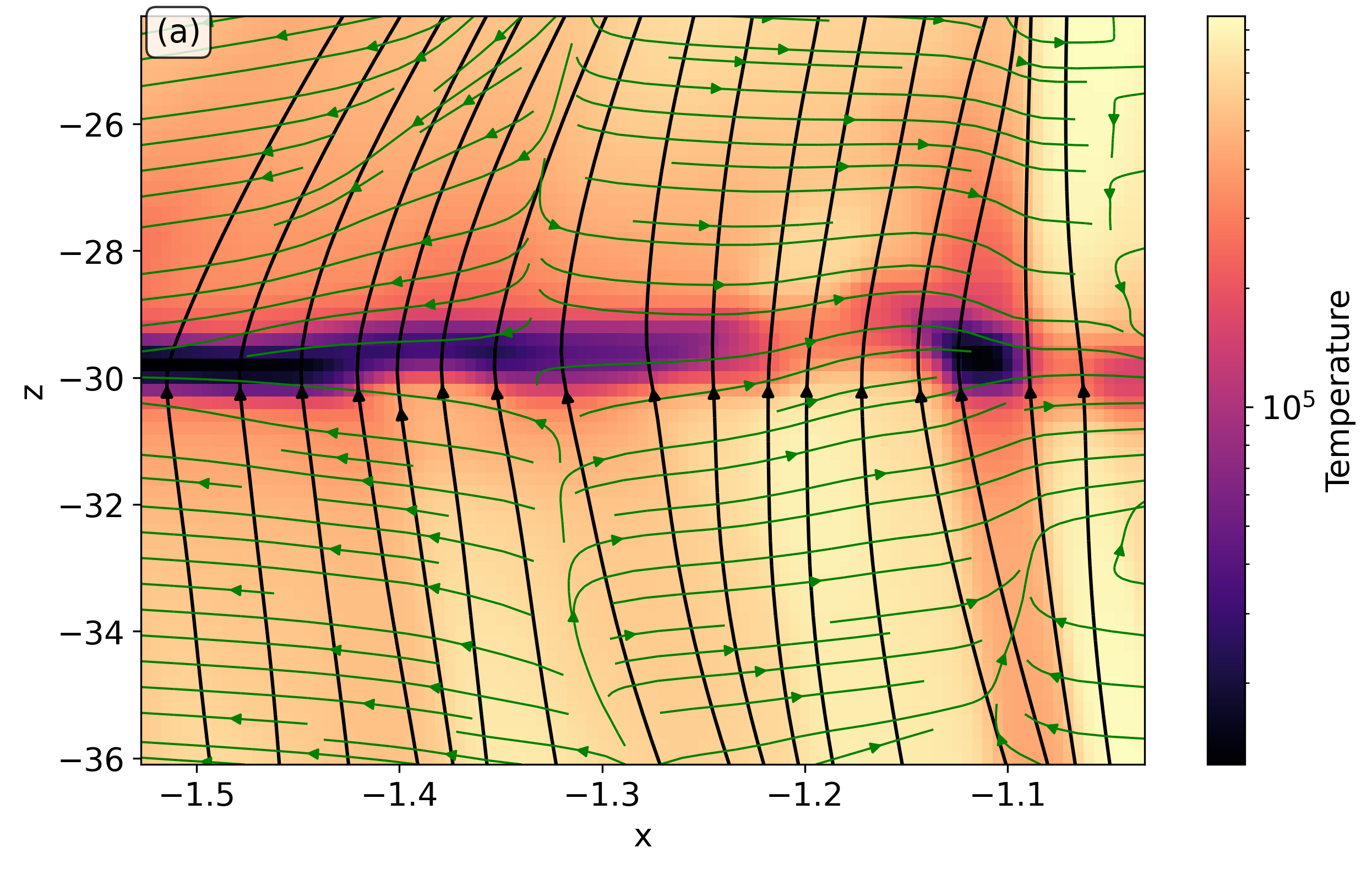} \\
    \includegraphics[width=0.95\linewidth]{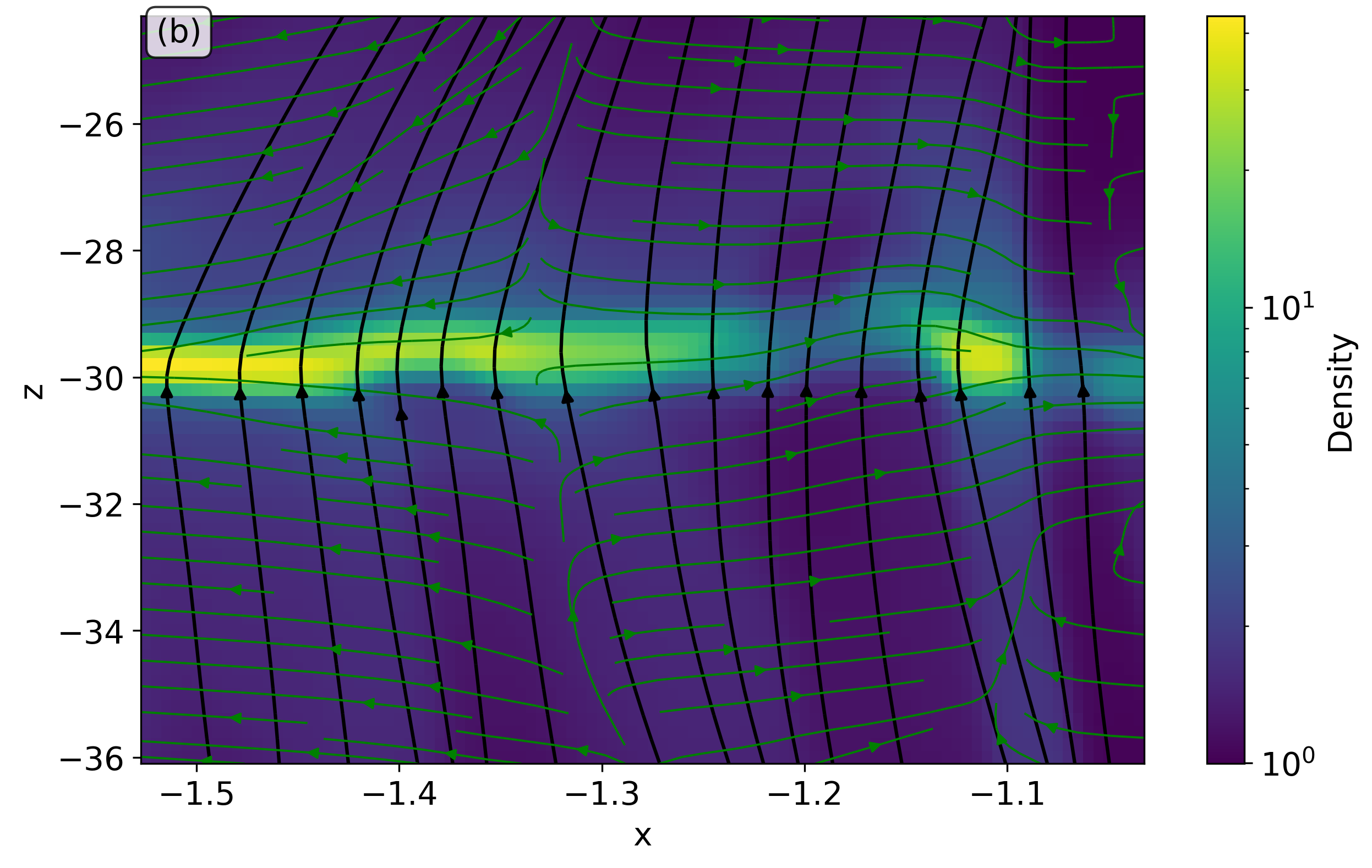} \\
    \includegraphics[width=0.95\linewidth]{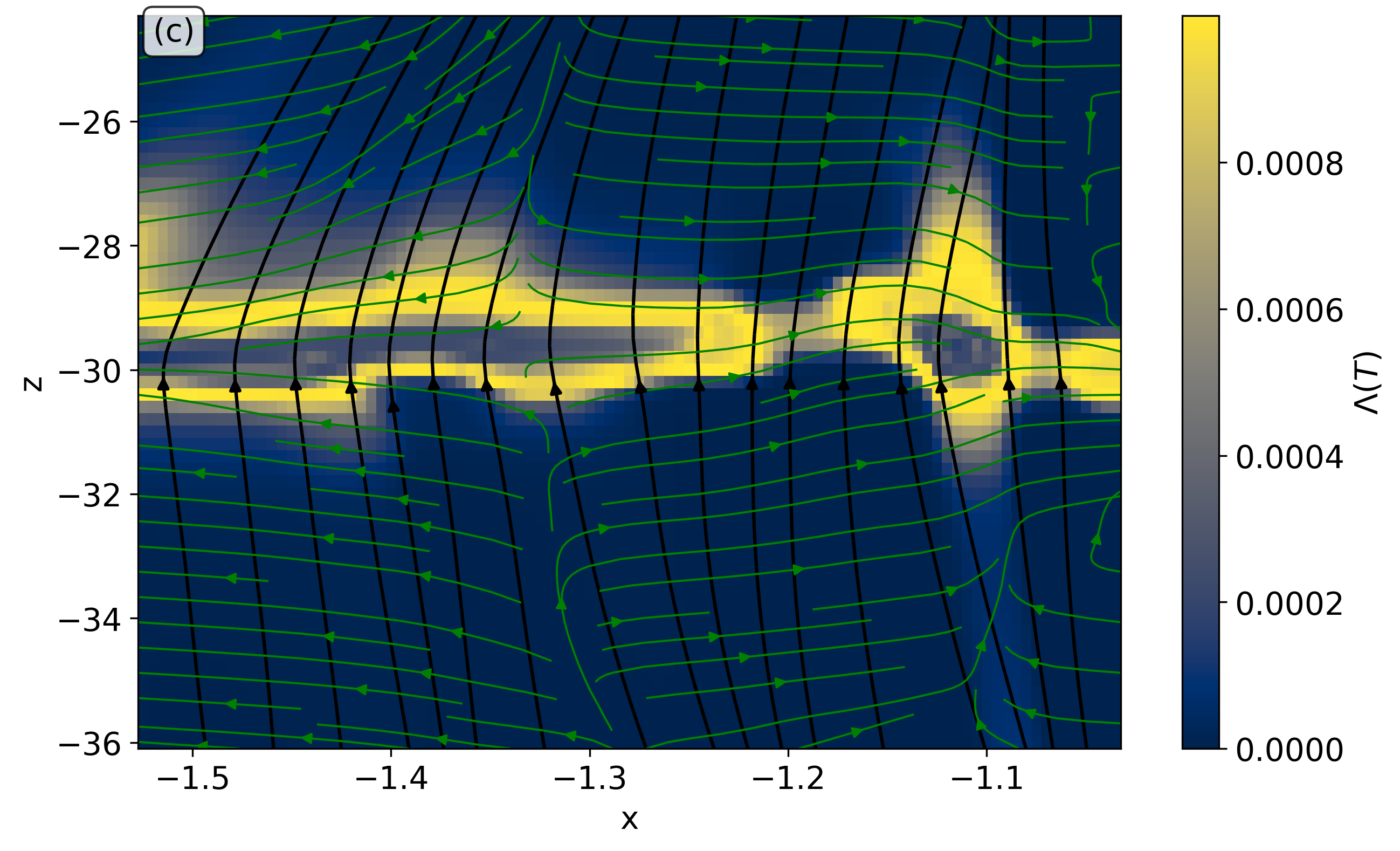} \\
    \includegraphics[width=0.95\linewidth]{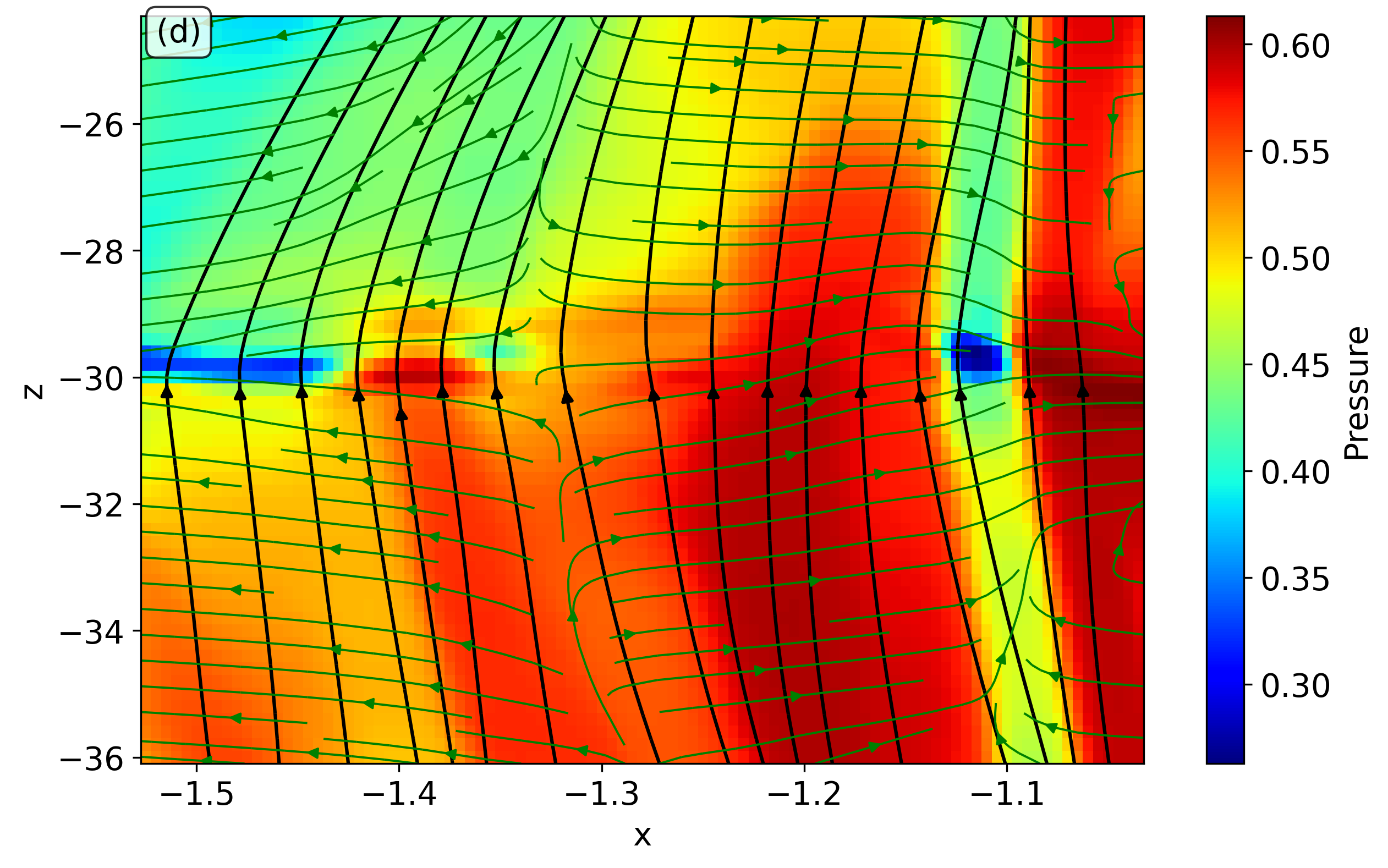}
    \caption{Properties of an extracted thermal instability at $y=-0.436$ and time $t=236$ showing the temperature (a), density (b), losses $\Lambda (T)$ (c), and pressure (d). The black fieldlines show the magnetic field. The green arrows show the perturbation velocity field in the plane ($\textbf{v}-\bar{\textbf{v}}$)}
    \label{fig:tecartoon}
\end{figure}

\begin{figure}
    \centering
    \includegraphics[width=0.95\linewidth]{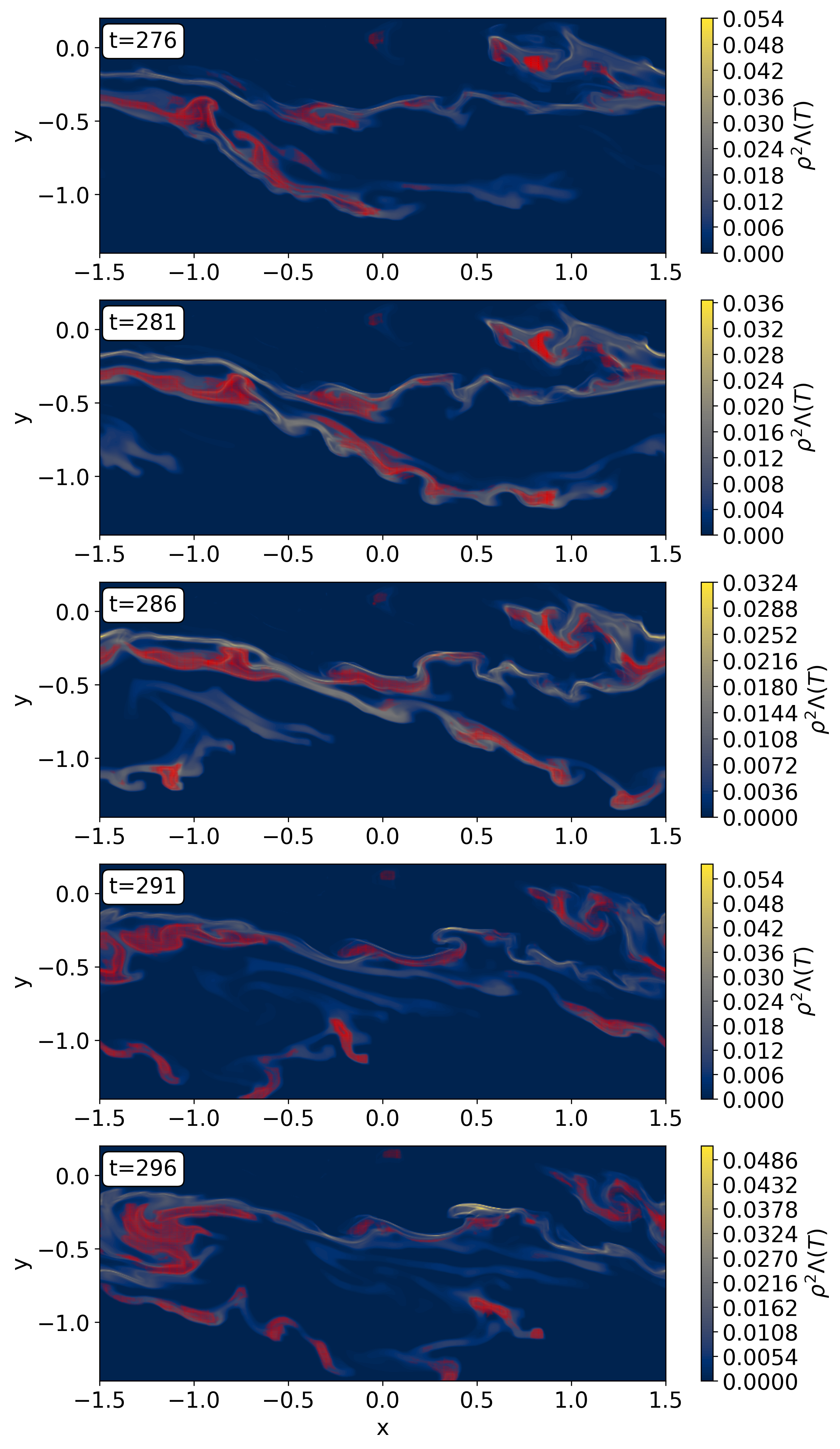}
    \caption{Slice of the mean thermal energy loss in a narrow plane ($35.9\leq z \leq 51.9$) around a thermal instability (colourmap). The red contour shows the thermal instabilities identified using the criteria outlined in Section \ref{sec:thermal_instabilities}.}
    \label{fig:tislice}
\end{figure}

To identify potential thermal instabilities, the following criteria are applied:
\begin{itemize}
    \item Converging flow alligned with the magnetic field: $\nabla \cdot \left( \textbf{v} \cdot \hat{\textbf{B}} \right) \hat{\textbf{B}}$, where $\hat{\textbf{B}}$ is the unit vector in the direction of the magnetic field. The thermal instability requires the inflow of material to support the instability, and thus this is a necessary condition. The gradient was calculated locally using second-order central difference.
    \item Non-zero radiative losses, specifically $\Lambda(T)>10^{-5}$, corresponding to the temperature range $9575\leq T \leq 560328$. The thermal instability must lose energy through radiation. This metric neglects the fully-cooled material which is no-longer thermally unstable.
\end{itemize}
Applying these metrics identifies a wealth of potential thermal instabilities, however, this detection mechanism is not exact. The detected features do possess the characteristic properties of thermal instabilities, as detailed below 

Figure \ref{fig:volume3D}d shows regions associated with losses $\Lambda(T)>10^{-5}$, and converging flow aligned with the magnetic field ($\nabla \cdot \left( \textbf{v} \cdot \hat{\textbf{B}} \right) \hat{\textbf{B}}$). As such, the features shown in Figure \ref{fig:volume3D}d are candidates for thermal instabilities, which arise as part of a multi-step process:
\begin{enumerate}
    \item material cools due to radiative losses, lowering the pressure
    \item The radiative losses are no longer efficient at that cooled temperature, however the pressure gradient drives flows that compress the region and raise the temperature
    \item Radiative losses become efficient again at the increased temperature, leading to cooling.
\end{enumerate} 
This process is a positive feedback cycle that repeats to form cool, dense regions with converging flow and moderate losses. The thermal instability can result in cool, dense condensations forming. Note that the condensation itself may be thermally stable (in that the energy lost due to radiation is negligible), whereas, by definition, the material undergoing thermal instability is not in a thermal equilibrium and subject to radiative losses.

In the simulation, the thermal instabilities manifest as long, narrow (in the $z-$direction) structures aligned predominantly in the $x-$direction, i.e, perpendicular to the magnetic field, as shown in Figure \ref{fig:volume3D}d. 
Their existence as thin structures perpendicular to the magnetic field is consistent with the literature \citep{Claes2020}. 
A close up of the properties around one of these structures is shown in Figure \ref{fig:tecartoon} for a slice in the $xz-$plane at $y=-0.436$. The cool dense material in the centre of the plot is aligned perpendicular to the magnetic field (shown by the black streamlines). The formation mechanism is that a small localised thermal instability forms which is then stretched in the $x-$direction by the flow (the shearing motions are in the $x-$direction), resulting in a long structure. The thermal instability is supported by flow along the magnetic field lines, which compresses the material and creates the positive feedback cycle needed for the thermal instability.

The perturbation velocity field (subtracting the mean flow in the window) is plotted as green field lines in Figure \ref{fig:tecartoon}. The perturbation velocity removes the influence of the bulk shearing motions in the $x-$direction. The magnetic field (black streamlines) is mostly aligned with the $z-$direction, however turbulent evolution has led to variations in the other direction. The temperature (Figure \ref{fig:tecartoon}a) and density (Figure \ref{fig:tecartoon}b) reveal a cool, dense tube, which has converging flow aligned with the magnetic field, and efficient cooling at the boundaries of the structure (Figure \ref{fig:tecartoon}c).

As stated in the introduction, thermal instabilities are broadly categorised as being isobaric (constant pressure), or isochoric (constant density) \citep{Field1965}. Now, in a turbulent mixing layer, purely isochoric or isobaric structures are unlikely to exist since the surrounding material is not isotropic. However, from the density (Figure \ref{fig:tecartoon}b) and pressure (Figure \ref{fig:tecartoon}d) colourmaps, the variations in density are significantly greater than the pressure variations in and around the thermal instability. As such, this is most likely an isobaric-type thermal instability, as one would expect for a solar-like thermal instability \citep{Field1965,Claes2019}.

Note that the resolution of the $z-$direction is significantly lower than the $x-$ direction due to the long domain required in the $z-$direction as a consequence of the Alfv\'en crossing time. As such, the apparent variation in the $z-$direction of the thermal instabilities is small, however this is a consequence of the large aspect ratio. From Figure \ref{fig:tecartoon}, the width across the thermal instability is $\approx1$ in the $z-$direction (estimated from the high density region), which is resolved by 5 grid cells. As such, the maximum density obtained within the thermal instabilities may be limited by the resolution and a higher resolution simulation could produce condensations with an even larger density. Note that mass is drawn into the thermal instabilities from beyond the 5 cell thickness, but this is the thinnest width that the thermal instability can collapse to in our simulation.

\subsection{Distribution of thermal instabilities}
\begin{figure}
    \centering
    \includegraphics[width=0.95\linewidth]{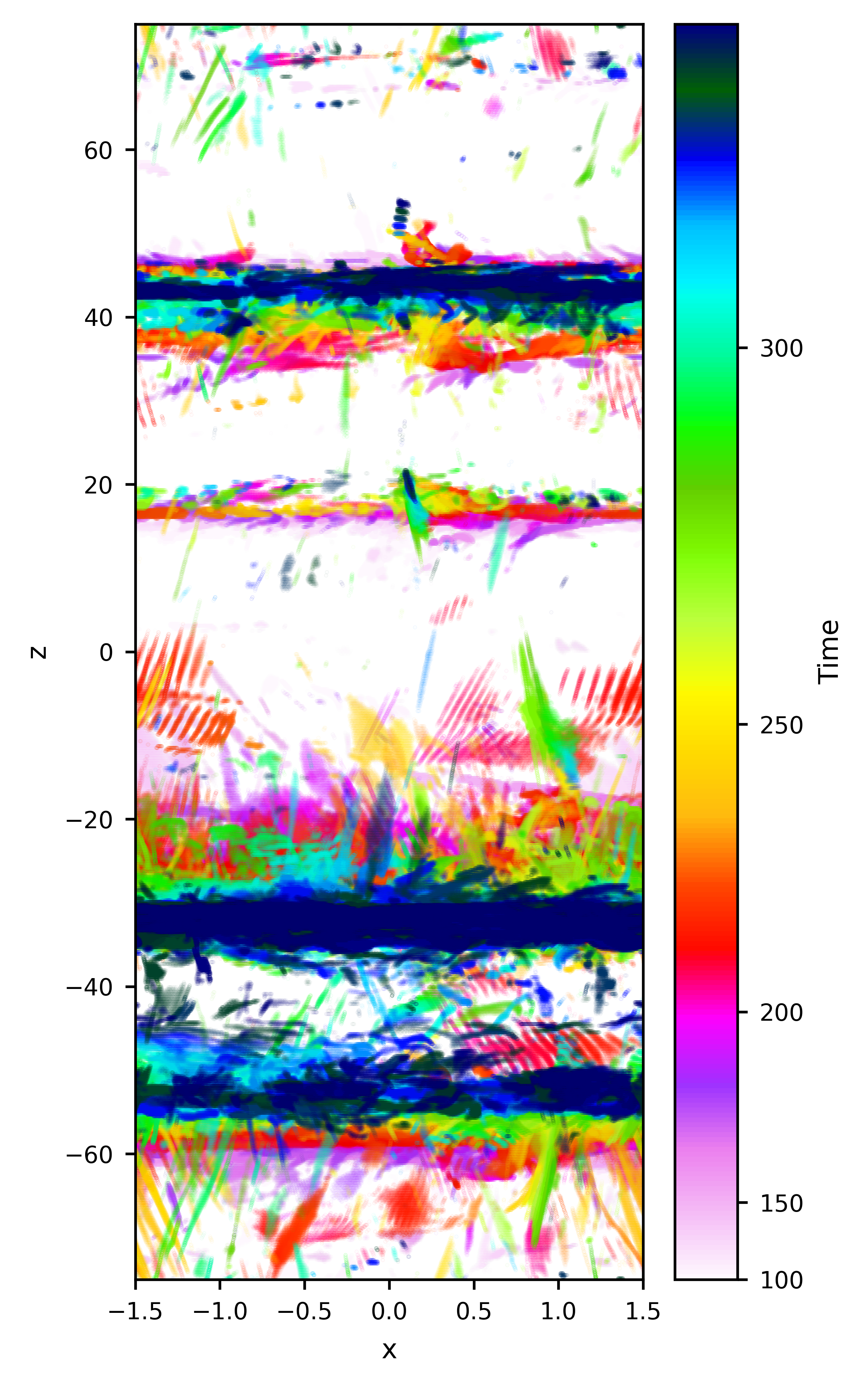}
    \caption{All detected thermal instabilities through time (colours), summed over the $y-$direction}
    \label{fig:ti_distribution}
\end{figure}

The turbulent motions of the mixing process can churn up the thermal instabilities causing them to fracture into smaller structures. The elongated structures seen in Figure \ref{fig:volume3D}d are examples of thermal instabilities that are stretched perpendicular to the magnetic field due to the velocity shear. However, these thermal instabilities are not monolithic structures, and are continuously evolving with the mixing layer. Figure \ref{fig:tislice} shows the thermal instabilities that occur in a narrow slice in the $z-$direction through time. The thermal instability structures are seen to continuously fragment and evolve with the dynamics in the plane. The fragmentation is driven by the underlying mixing dynamics agitating the thermal instabilities, as seen in Figure \ref{fig:tislice}.  In \cite{Claes2020}, fragmentation of thermal instabilities arises as a result of thin-shell instabilities, however, here the underlying mixing is responsible. 

{Many small-scale thermal instabilities form throughout the simulation. However, these are typically short-lived, or become absorbed into the large-scale thermal instabilities. Figure \ref{fig:ti_distribution} shows all detected thermal instabilities in the time range $t=\{0,350\}$ through the simulation, coloured by the simulation time. The most prominent features are the long-lived structures, and many smaller, short-lived features form sporadically through the domain. The shearing flow in the $x-$direction stretches the thermal instability structures creating these band-like structures. Further instabilities appear to form more readily in the areas surrounding these bands.}


\subsection{$x$-averaged mixing layer}

\begin{figure*}
    \centering
    \includegraphics[width=0.99\linewidth]{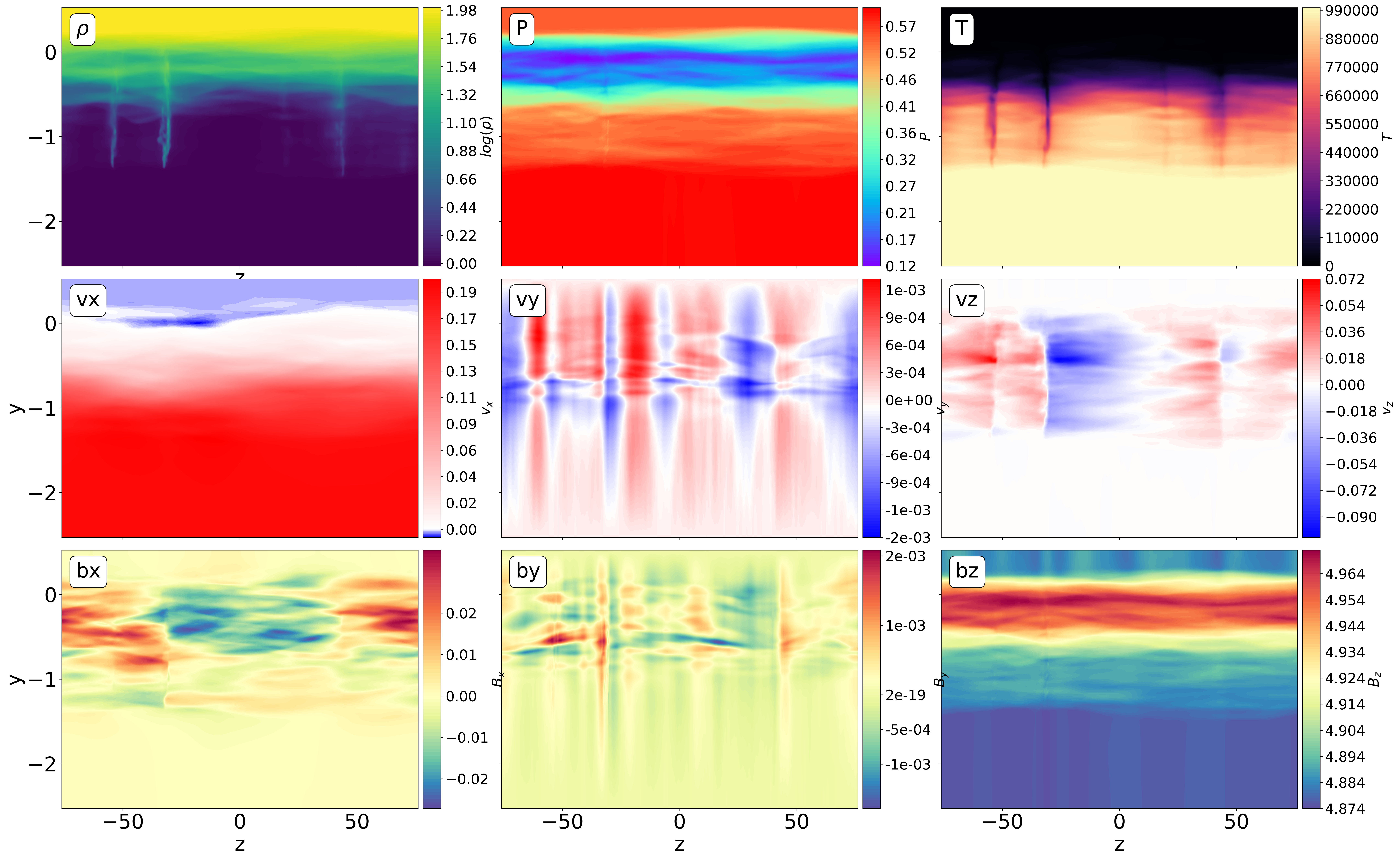}
    \caption{$x-$averaged quantities at time $t=296$ showing density (top left), gas pressure (top centre), temperature (top right), $v_x$ (middle left), $v_y$ (middle centre), $v_z$ (middle right), $B_x$ (lower left), $B_y$ (lower centre), $B_z$ (lower right).}
    \label{fig:xave}
\end{figure*}

\begin{figure}
    \centering
    \includegraphics[width=0.95\linewidth]{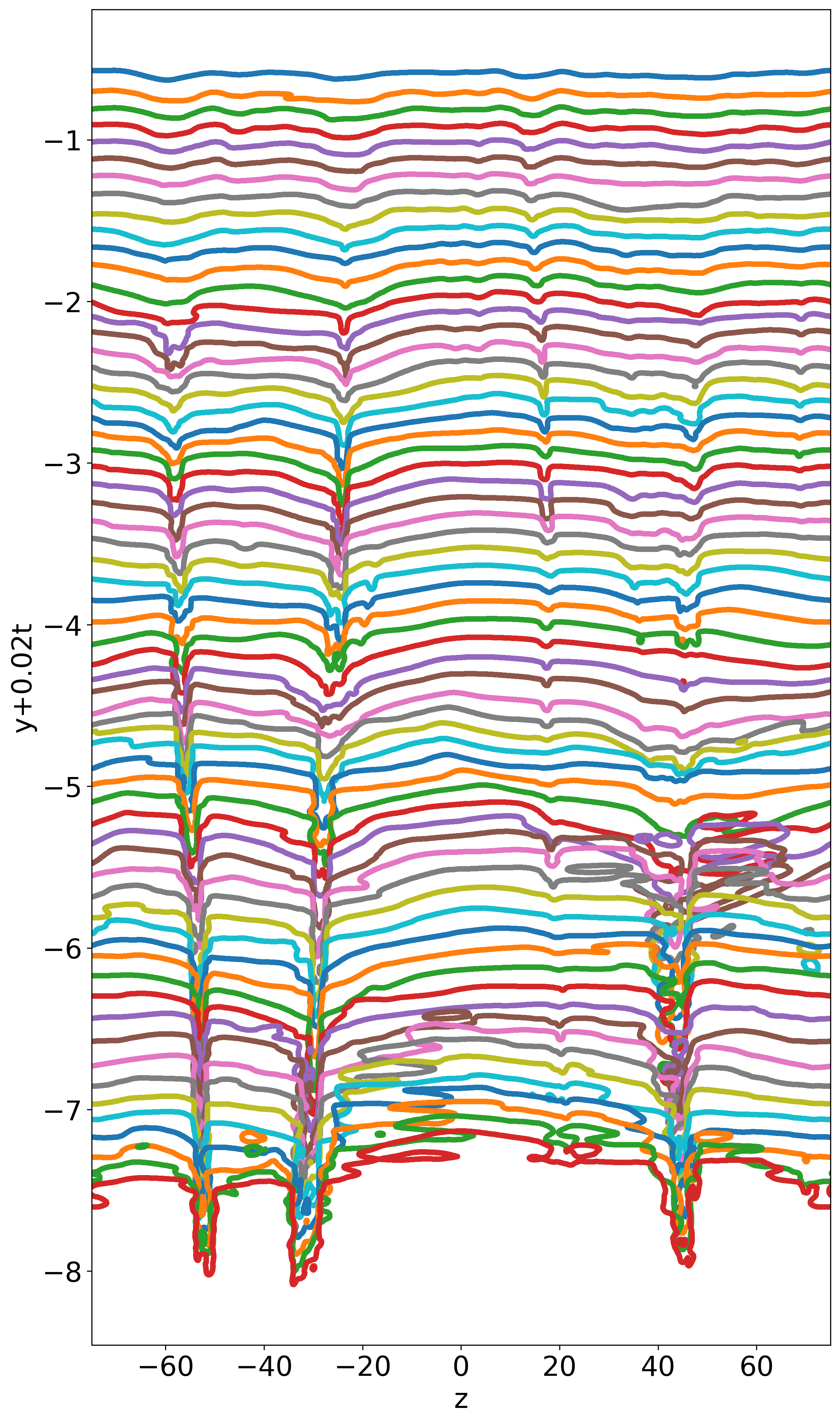}
    \caption{Lower bound of the mixing layer based on the $x-$averaged density for the time range $25\leq t \leq 336$, shifted by $0.02 t$. The lower bound is determined based on the $x-$averaged density $\rho_x=2$.}
    \label{fig:mixingwidth}
\end{figure}

As can be seen in Figure \ref{fig:volume3D}d, the thermal instabilities are aligned perpendicular to the magnetic field, along the $x-$axis, spanning nearly the full extent of the domain. As such, $x-$averaged properties allow the mean quantities of the mixing layer and thermal instabilities to be studied to gain insight into the overall effect the large-scale thermal instabilities have on the dynamics. An $x-$averaged quantity is defined as
\begin{gather}
    \bar{\xi}_x (y,z)=\frac{\int^{x=1.5}_{x=-1.5} \xi(x,y,z)dx}{\int^{x=1.5}_{x=-1.5}dx}
\end{gather}
for an arbitrary quantity $\xi$.

The $x-$averaged properties are shown in Figure \ref{fig:xave}. The large thermal instabilities are clearly visible in the density and temperature panels and manifest as elongated cool, dense structures that permeate into the low-density region. Note that the thermal instabilities exist within the mixing layer, as evidenced by the other panels in Figure \ref{fig:xave}. However, the lower part of the mixing layer is low density and much of the mass is absorbed into the thermal instabilities, indicating that the density may not be a good indicator of the mixing layer width since the height-averaged mass is influenced by the thermal instabilities.  

The inflow that supports the thermal instabilities comes predominantly along the $z-$direction, i.e., along the magnetic field, as can be seen by the converging flow in Figure \ref{fig:xave}. Whilst the largest magnetic field component is the $B_z$, transverse magnetic field is generated throughout the mixing layer and jumps in the $B_x,B_y$ components are present. The converging flow in the thermal instabilities brings in the surrounding magnetic field variations and effectively compresses them. As such, the local variation in transverse magnetic field are larger at thermal instabilities than in the surrounding plasma.

The gas pressure is noticeably lower in the mixing layer than outside the mixing layer. Most of this pressure drop is restored by the magnetic pressure, in particular, the $B_z$ component, which has a relatively small increase in the mixing layer. Initially, the system has a plasma-$\beta=0.05$, i.e., the magnetic pressure is far stronger than the gas pressure. As such, a small increase in $B_z$ is sufficient to balance the pressure. Whilst there are some traces of the thermal instabilities present in the pressure field, it is relatively small variation, lending support to the thermal instabilities being primarily isobaric-like.

In Figure \ref{fig:mixingwidth}, the thermal instabilities are seen to attract each other. As the plasma cools, the pressure reduces leading to pressure-driven flows which replenish the pressure and feed the instability. The inflow has the effect of reducing the pressure slightly outside the thermal instability which causes the thermal instabilities to attract. The attraction of thermal instabilities has been extensively studied in the context of cloud coalescence \citep{Waters2019}. 

\subsection{Cool mass evolution}

\begin{figure}
    \centering
    \includegraphics[width=0.95\linewidth]{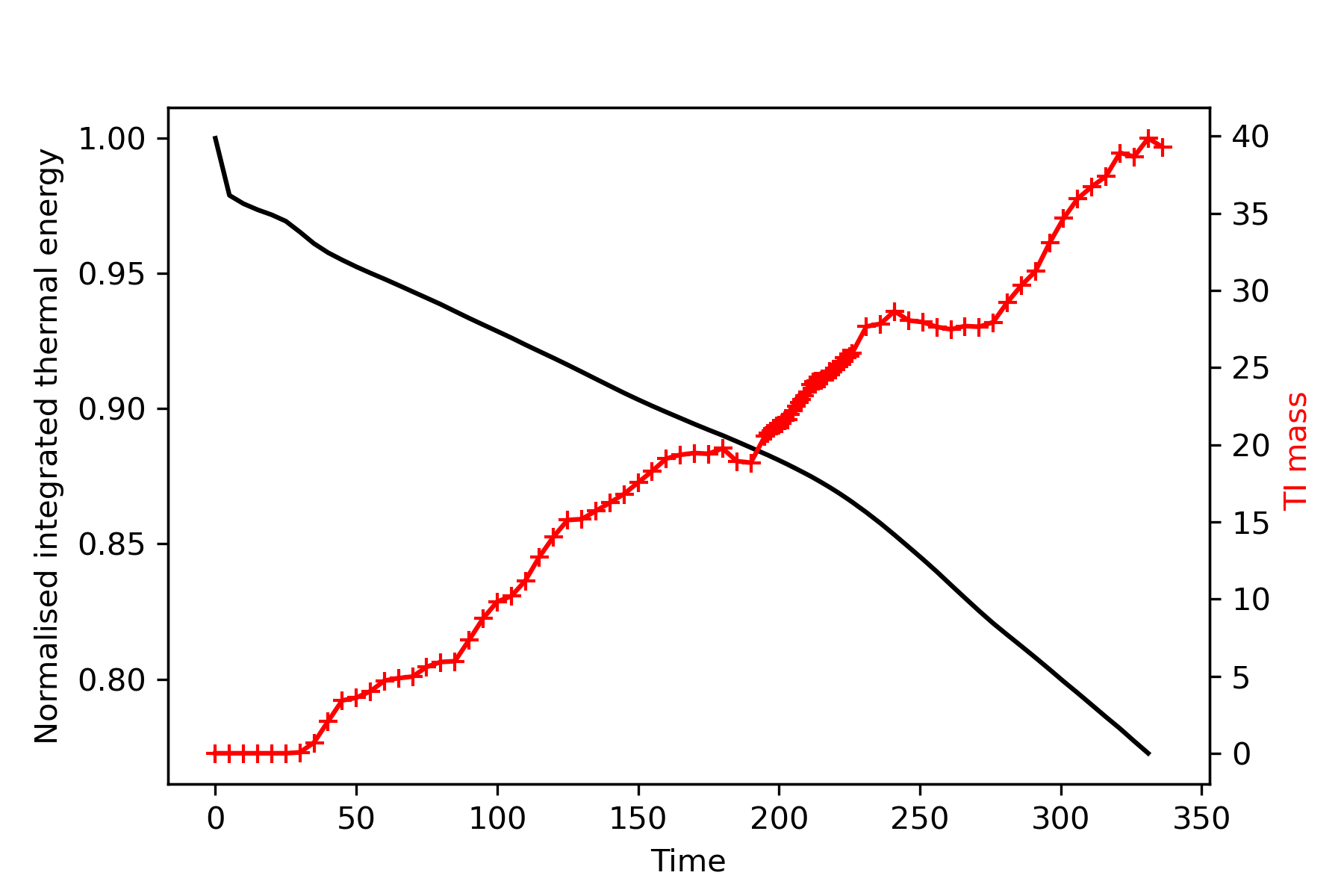}
    \caption{Integrated thermal energy over the domain through time, normalised by the value at $t=0$ (black). Red line shows the mass contained within thermal instabilities through time.}
    \label{fig:energycomp}
\end{figure}

\begin{figure}
    \centering
    \includegraphics[width=0.95\linewidth]{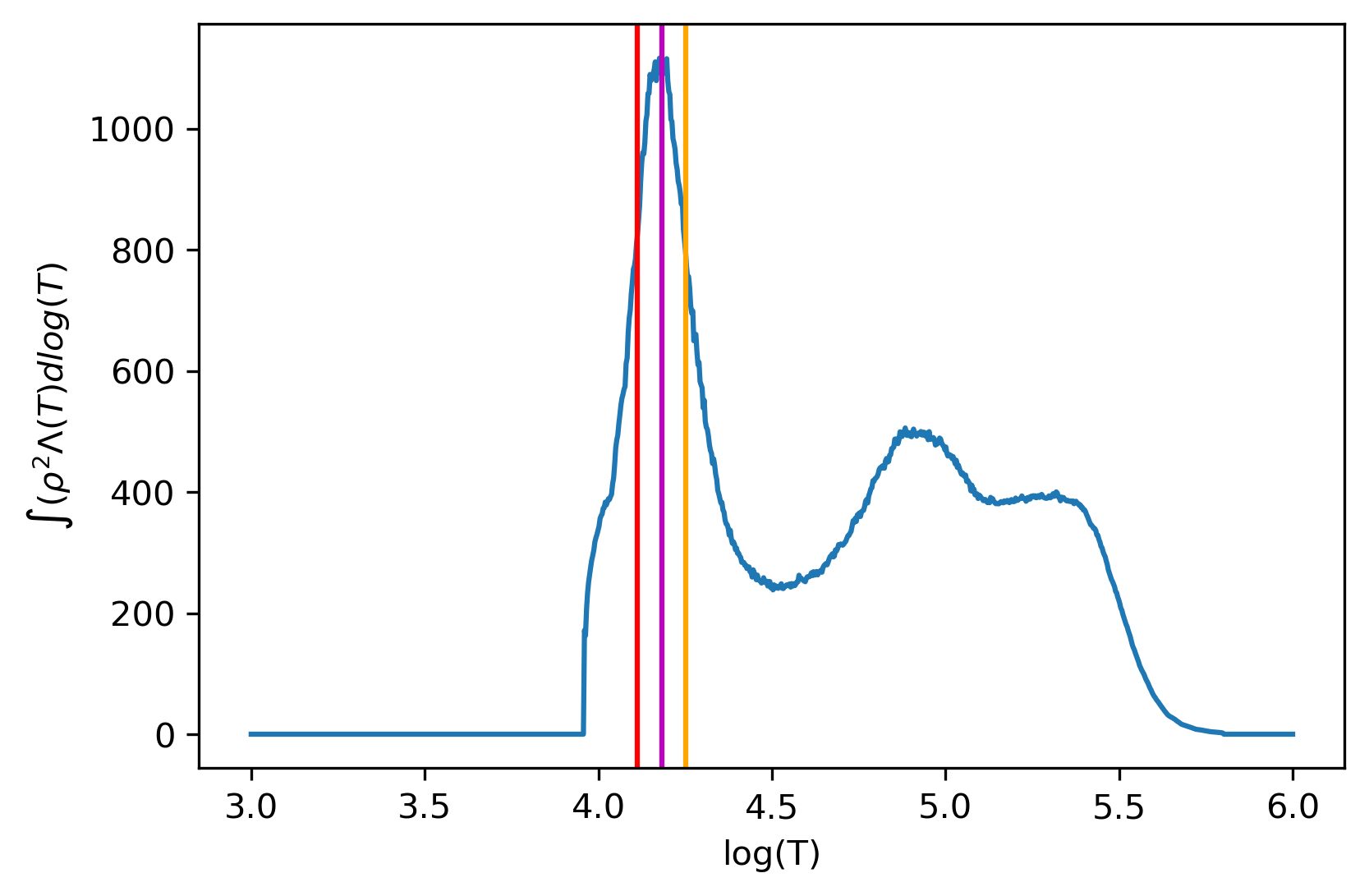}
    \caption{Integrated cooling rate ($\int \rho^2 \Lambda(T) dT$) over a narrow temperature band in log space. 1000 bands are specified in the range $\log(T)=3,6$. The vertical lines show the temperatures $T_{\rm cut}$ that have 90, 80, 70 \% of the total losses above them, as defined by Equation \protect\ref{eqn:tcut}.}
    \label{fig:masslossdist}
\end{figure}

\begin{figure}
    \centering
    \includegraphics[width=0.95\linewidth]{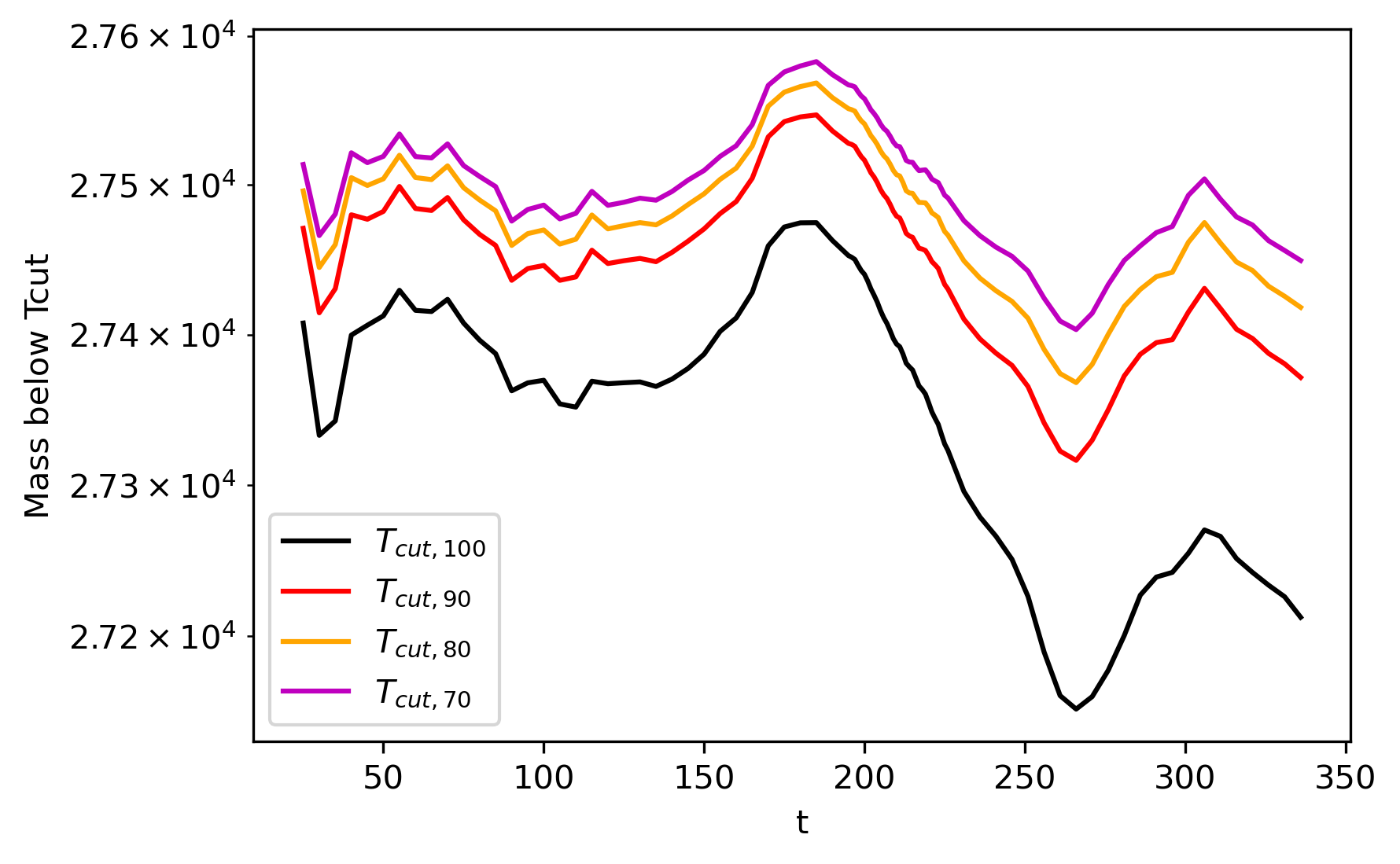}
    \caption{Total mass below $T_{\rm cut}$ through the simulation for $T_{\rm cut}=$ $T_{\rm cut,100}$ (black), $T_{\rm cut,90}$ (red), $T_{\rm cut,80}$ (magenta), $T_{\rm cut,70}$ (orange). }
    \label{fig:massbelowtcut}
\end{figure}

The formation of cross-field thermal instabilities results in cool, dense material as a result of radiative losses in the Kelvin-Helmholtz instability. The mass contained within thermal instabilities increases with time, as shown in Figure \ref{fig:energycomp}. In cloud-crushing simulations with strong radiative losses, thermal instabilities can lead to the formation of additional cool material due to the thermal instabilities, such that the total cool mass can exceed the initial cool mass \citep{Gronke2018}.

The total cool mass in our simulation depends on the definition of `cool'. Following \cite{Hillier2019}, the cool temperature can be defined as the temperature at which radiative losses are no-longer efficient, {i.e., $T_{\rm cool}=9575{\rm K}$. This is a hard limit of cooling since there is no non-adiabatic mechanism in the simulation to cool the plasma below this. Note that adiabatic expansion can further reduce the temperature, but this is not a cooling mechanism.} However, the hard limit of cooling neglects material that is in a cooling state, but has not yet fully cooled, which may represent a significant mass in a turbulent simulation. To include this material in the calculation of the total cool mass, an effective cut-off temperature $T_{\rm cut}$ must be defined from the data.

Figure \ref{fig:masslossdist} shows the integrated energy loss as a function of temperature. Following \cite{Hillier2019}, the energy loss rate can be estimated using a temperature at which radiative losses are no-longer effective, Temperature is banded in log space to create a histogram that shows how much cooling is occurring in each narrow temperature band. Whilst the peak of $\Lambda(T)$ occurs at around $\log(T)=5.5$ (see figure \ref{fig:lossfunc}), the loss rate depends on the density squared and, as such, cool material entering the mixing layer is responsible for a far larger energy loss, peaking around $\log(T)=4.2$, as shown in Figure \ref{fig:masslossdist}. The vertical lines in Figure \ref{fig:masslossdist} indicate the temperature above which $90\%$ (red), $80\%$ (magenta) and $70\%$ (orange) of the total losses occur, i.e., the red line marks the temperature $T_{\rm cut,90}$ where
\begin{gather} \label{eqn:tcut}
   \int_{T_{\rm cut,90}}^{T_{\rm max}} \rho^2 \Lambda(T) \,dV = 0.9 \int_{0}^{T_{\rm max}} \rho^2 \Lambda(T) \,dV
\end{gather} 
i.e., where 90\% of all losses occur above $T_{\rm cut,90}$, and similarly for $T_{\rm cut,80},T_{\rm cut,70}$. The cut-off temperatures are determined from a single, late-stage simulation when the mixing layer is turbulent and a wide range of temperatures (and loss rates) are present.

The integrated mass below different levels of $T_{\rm cut}$ is shown through time in Figure \ref{fig:massbelowtcut}. All values of $T_{\rm cut}$ show roughly the same behaviour and there is no monotonic trend in the quantity of mass below $T_{\rm cut}$. In the previous study of \cite{Hillier2023}, the mass below $T_{\rm cut}$ monotonically increased. The difference here is likely due to the more moderate losses included in the present simulation. Appendix \ref{sec:appMassBelowTcut} shows the mass below $T_{\rm cut}$ in the 2D simulations of \cite{Hillier2023} for a range of loss rates. Whilst there is no significant change in the total mass below $T_{\rm cut}$, the value being reasonable steady implies that cool material is being replenished due to the radiative losses. 





It is known that for strong losses, the mass below $T_{\rm cut}$ increases, whereas shearing instabilities with long radiative timescales (i.e., weak losses) show a decrease in cool mass \citep{Gronke2018,Hillier2023}. Here, the likely scenario is that the losses are occurring on an intermediate timescale, relative to the dynamics, thus the total cool mass in the system is roughly constant. The radiative losses are sufficiently strong to substantially reduce the total thermal energy of the system (as discussed later in Section \ref{sec:energyloss}) and form localised cool, dense structures through thermal instabilities. However, the net mass of cool material does not substantially change through the simulation. 



\subsection{Integrated energy loss} \label{sec:energyloss}

The integrated thermal energy through time is shown in Figure \ref{fig:energycomp}. Initially, there is a sharp drop in thermal energy due to the finite losses at the initial temperature of the lower domain. The cooling associated with this is small (as shown by the cooling curve in Figure \ref{fig:lossfunc}), however occurs over a large area and thus the losses are apparent in the integrated thermal energy plot. 

Through time, the thermal energy of the system decreases due to the radiative losses. The initial plasma either side of the interface is thermally stable (in the sense that the radiative losses are effectively zero). However, the mixed material forms at temperatures that have efficient radiative losses which removes thermal energy from the system. Turbulent heating exists due to the finite numerical diffusion, however the radiative losses are far greater than the turbulent heating. 2D simulations have shown that even when radiative losses occur on long timescales, they are still efficient at removing thermal energy from the system \cite{Hillier2023}. The rate at which thermal energy is lost is roughly constant until time $t \approx 200$, after which the loss rate increases and energy is removed from the system more efficiently. A change in the rate of the thermal energy loss was also seen in the 2D simulations of \citet{Hillier2023}. 


\subsubsection{Energy loss rate associated with thermal instabilities} 

\begin{figure}
    \centering
    \includegraphics[width=0.95\linewidth]{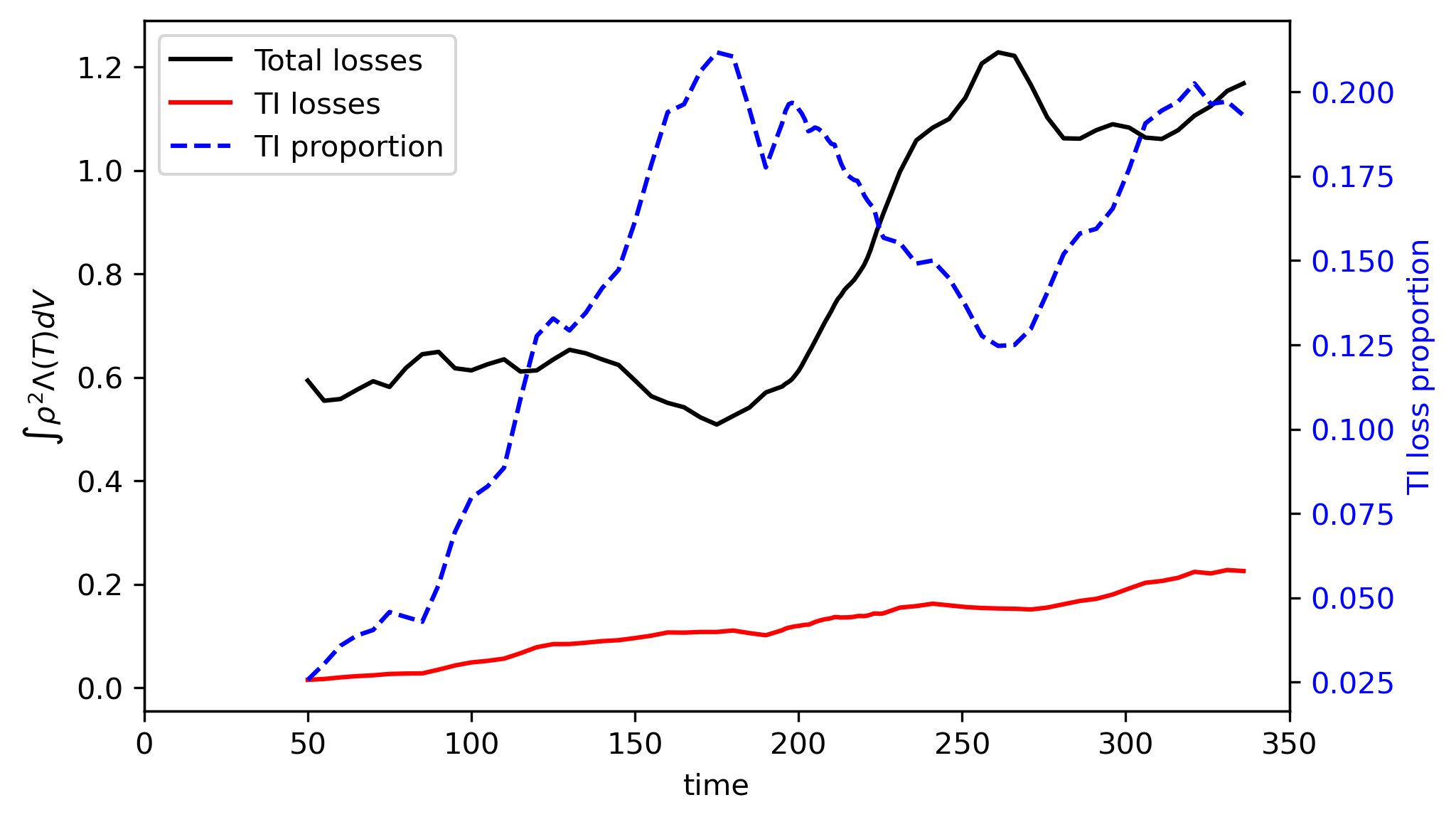}
    \caption{Integrated energy loss $\rho ^2 \Lambda(T)$ in the total domain (black) and the identified thermal instabilities alone (red). The blue line shows the proportion of thermal energy loss due to thermal instabilities and is associated with the right axis.}
    \label{fig:tiloss}
\end{figure}


Radiative losses occur within the mixing layer as well as the thermal instabilities. Figure \ref{fig:tiloss} shows the integrated thermal energy loss due to radiation $\int \rho^2 \Lambda(T) dV$ evaluated over the entire domain, and the thermal instabilities. At early times ($t<100$), thermal instabilities are responsible for a small amount of the total losses ($<10\%$). At early time, the mass contained within thermal instabilities is small and thus one would not expect them to have a significant impact on the total thermal energy loss.

As the system evolves, the thermal instabilities grow and contain more total mass, as shown by the red line in Figure \ref{fig:energycomp}. Since thermal instabilities are sites of a strong loss cycle, the energy loss due to thermal instabilities becomes a greater proportion of the total energy loss ($\approx20\%$), as shown in Figure \ref{fig:tiloss}. The peak proportion of energy loss occurs around $t=170$. Following this, the total losses increases at around $t=200$ (coinciding with the change in energy loss rate, Figure \ref{fig:energycomp}), with the proportion of losses explained by thermal instabilities accounting for approximately $15-20\%$ of all losses.


\section{Discussion}

\subsection{How 3D is the turbulence?} \label{sec:3Dness}

A common discussion for 3D MHD turbulence is the extent to which the system truly exhibits 3D turbulence, or is effectively a series of 2.5D slabs. The volume plots in Figure \ref{fig:volume3D} indicate that there is 3D behaviour but a more formal analysis is beneficial. The main distinction between the fully 3D case and a series of 2.5D slabs is whether the lengthscales of the system are sufficient that independent structures can grow in each direction. All turbulence setups have unique conditions and thus specific metrics for estimating the 3D-ness of the system vary. For forced 3D turbulence with a guide field,
\citet{Alexakis2011} give a criteria where 3D simulations can {become} 2D-like under the condition that
\begin{gather}
    \frac{|v_k|k_\perp}{V_{A0}k_\parallel} \ll \frac{1}{k_\perp L_\perp} \frac{1}{\sqrt{k_\parallel L_\parallel}} \label{eqn:3dness}
\end{gather}
where $k_\parallel$ is the wavenumber in the direction of the field, and $k_\perp$ is perpendicular to the field. In our setup, the reference speed $v_k$ and Alfv\'en Mach number $V_{A0}$ have multiple possible choices.

To consider the effect on the larger-scale evolution of the mixing layer, one can choose properties of the mixing layer itself. The height of the mixing layer is calculated as when the mean density in the $xz$-plane is between 2 and 98, giving the $y-$range of the mixing layer as $\left[-0.717, 0.278\right]$. As such, the wavenumber of the mixing layer is $k_\perp\approx2\pi/1$(wavenumber of the mixing layer thickness). The perpendicular wavenumber is the full extgent of the $z-$domain, i.e., $k_\parallel=2\pi/150$. For the velocity and Alfv\'en speed, we will use the RMS values of the mixed material, giving $v_{RMS}\approx0.052$, and $V_{A,RMS}\approx 2.65$. All quantities are calculated at time $t=336$. 
Substituting these into the inequality gives
\begin{gather}
    \frac{|v_{RMS}|k_\perp}{V_{A,RMS}k_\parallel} \approx 2.94 \\
    \frac{1}{k_\perp L_\perp} \frac{1}{\sqrt{k_\parallel L_\parallel}} \approx 0.0635 
\end{gather}
As such, one would expect that the 3D modes are able to grow on the scales of the mixing layer. 




\subsection{Aspect ratio} \label{Sec:aspect}

The simulation presented in this study has a large aspect ratio of 40:1, where the grid cells are elongated in the $z-$direction, which is aligned with the magnetic field. This aspect ratio was chosen to account for the strong Alfv\'en speed in the $z-$direction requiring a large domain to exhibit 3D structure. Since large aspect ratios can affect the dynamics, it is necessary to consider the effect of our aspect ratio. The effect at the grid-cell level can be considered by applying the same criterion as above, however, with the wavenumbers determined by the grid cell itself, i.e., $k_\parallel=1/\Delta z=5$ (wavenumber of a z-grid cell), $k_\perp=1/\Delta x=200$ (wavenumber of an x-grid cell). The same values of $v_{RMS}\approx0.052$, and $V_{A,RMS}\approx 2.65$ are used to study the mixing layer itself. Using these quantities, 
\begin{gather}
    \frac{|v_{RMS}|k_\perp}{V_{A,RMS}k_\parallel} \approx 0.785 \\
\end{gather}
As such, the there is approximate coherency of Alfv\'en waves at grid scale. Even though the aspect ratio is very high, the anisotropy of wave speeds means that information travels at approximately the same rate in all directions on our grid. Therefore, the high aspect ratio is not expected to have a significant effect on the results.

\subsection{The role of mixing-induced cooling in forming condensations in the solar corona}

The key result of this paper is that mixing-induced cooling in 3D KHI can lead to the formation of not only cool structures \citep[as seen in][]{Hillier2023} but also cool, dense structures as a result of condensations forming in the mixing layer as a consequence of the thermal instability. This instability has been studied extensively in the context of prominence formation and coronal rain \citep[e.g.][]{Antolin2020}, and here we have shown it could also be active in the PCTR as a consequence of mixing, which is an important consequence for understanding the role of mixing-induced cooling in the solar atmosphere.



The thermal instabilities in this simulation are complex structures that evolve over time. As such, one would expect that the mass of thermal instability induced condensations is variable with time. Here, the thermal instabilities are seen to form perpendicular to the magnetic field. It has been shown that multiple thermal instabilities can occur along a single field-line in 1D prominence models \citep{Terradas2021A&A...653A..95T}
, which may be related to the formation of condensation in the solar atmosphere (e.g., prominences and coronal rain). 

An important caveat with this simulation is that the domain itself provides a lengthscale, and coupled with the strong mean-field shear flow and periodic boundaries, acts to stretch the thermal instabilities into long, thin structures aligned with the direction of shear (i.e., the $x-$direction). In a more realistic case, a turbulent shear flow and a larger domain may allow a more sporadic thermal instability structure in prominence-corona mixing layers. As such, the elongated thermal instability structures are most-likely to form in falling prominence thread type scenarios, where the shearing motion is determined by gravity, and hence directed. 

Our simulations show that the self-consistently generated thermal instabilities within the Kelvin-Helmholtz instability subject to radiative losses can maintain the total amount of cool mass. If radiative losses occurred on shorter timescales, one would expect that the amount of cool mass would increase more, up to a limit determined by the mixing \citep{Hillier2023}. The cooling rate used in this paper is an underestimate of the expected solar cooling rate for large-scale structures, and the simulations presented is far from the mixing-limit. As such, one would expect that thermal instabilities form readily in prominence-corona-like mixing scenarios.

\subsection{Turbulent thermal conduction}\label{turbTC}

In this paper, thermal conduction has been neglected which would be expected to have significant influence on the development of the thermal instability. The lengthscales that are unstable to thermal instability are given by the Field's length \citep{Field1965,Koyama2004ApJ...602L..25K}
\begin{equation}
    \lambda_{\rm F}=\sqrt{\frac{T_0 \kappa}{\rho_0 \Lambda(T_0)}},
\end{equation}
which without thermal conduction (i.e. $\kappa=0$) this would be zero. However, Figure \ref{fig:xave} shows separation between the cool dense condensations of $20$ to $60$ length units. This leads to the question, \textit{what creates the large separation between the condensations in this simulation?}

To answer this, we look at the role of turbulence to act as a transport term that creates an effective thermal conduction. Dimensionally the turbulent thermal conduction will be $\kappa_{\rm turb}\approx v_{\rm RMS}L_{\rm turb}$ with $v_{\rm RMS}$ the root-mean-squared turbulent velocity and $L_{\rm turb}$ is the characteristic lengthscale of the turbulence. From the mixing model of \citet{Hillier2019} and \citet{Hillier2023} we can state that
\begin{equation}
    v_{\rm RMS} = \frac{1}{2}\frac{(\rho_1\rho_2)^{1/4}}{\sqrt{\rho_1}+\sqrt{\rho_2}}\Delta V {=0.02878}
\end{equation}
{with $\rho_1=1, \rho_2=100, \Delta V=0.2$. Note that compared to the formulas in \citet{Hillier2019},} an extra factor of $1/\sqrt{2}$ is included to take into account the expectation that approximately half the energy of the turbulent fluctuations is in the magnetic field. We approximate $L_{\rm turb}$ by the mixing layer width $W=C_1v_{\rm RMS}t$ where for 3D mixing we take $C_1=0.3$ \citep{Baltzer2020JFM...900A..16B}. 

Taking this approximation for the turbulent thermal conduction our Field's length becomes
\begin{equation}
    \lambda_{\rm F - TURB}=\sqrt{\frac{T_{MIX}}{10^6}\frac{ \kappa_{\rm turb}}{\rho_{\rm MIX} \Lambda(T_{\rm MIX})}}\approx\sqrt{C_1\frac{T_{MIX}}{10^6}\frac{v_{\rm RMS}^2 }{\rho_{\rm MIX} \Lambda(T_{\rm MIX})}t},
\end{equation}
where the subscript $\rm MIX$ implies the characteristic value for these quantities in the mixing layer based on the theory of \citet{Hillier2019}{ Here, the normalisation is based on the coronal sound speed, hence a factor of $10^6$ is required to normalise the temperature. This gives the mixing quantities as:
\begin{gather}
    \rho_{MIX}=\sqrt{\rho_1 \rho_2}=10\\
    \frac{T_{MIX}}{10^6}=\sqrt{T_1 T_2} = 0.1
\end{gather}
and $\Lambda(T_{MIX})=10^{-3}$, i.e., the value at the maximum of the cooling curve. 
This gives the lengthscale of $\lambda_{\rm F - TURB}\approx 0.05\sqrt{t}$, which at time $t=300$ gives $\lambda_{\rm F - TURB}\approx0.85$ (which does not explain the distance between the condensations seen in Figure \ref{fig:xave}).}

{There is an} important qualification to this estimate. The turbulent transport is naturally anisotropic in this system due to the Alfv\'{e}n waves creating coherency along the strong magnetic field \citep{Hillier2019, Russell2025ApJ...980..186R}. This occurs because the vortical motions perpendicular to the magnetic field created by the KHI are coupled to the parallel direction by the Alfv\'{e}n waves these motions produce. 
As such, the natural lengthscale along the magnetic field ($L_{\rm turb}$) created by the largest vortical motions in the mixing layer would approximately be $L_{\rm turb}\approx W V_A/v_{\rm RMS}$ (where $V_A$ is the characteristic Alfv\'{e}n speed of the mixing layer and $W$ is the width in the y direction of the turbulent layer). This gives along the magnetic field $\lambda_{\rm F - TURB} \approx 0.37 \sqrt{t}$ or $\sim6.1$ at $t=300$. This gives a greater guidance as to why we see separations between condensations of a few tens of units, because those below $6$ units are completely stable. Note this parallel scaling is presented more as an indicative argument where greater accuracy could be found through detailed analysis of the turbulent structures in the mixing layer.

Putting this into context for the solar atmosphere, it is known that thermal conduction can suppress the thermal instability in the solar corona over significant lengthscales \citep[$\approx 2800$km, assuming a number density of $10^{9.5}\ \mbox{cm}^{-3}$ and temperature $T\approx 10^{5.8}$ K, ][]{Antolin2020}. However, in the mixing layer cool material and coronal material the characteristic temperature is approximately an order of magnitude smaller but both the characteristic density and cooling function are an order of magnitude larger. Using Spitzer-Harm thermal conduction $\kappa \propto T^{5/2}$ which implies the Field's length is approximately $10^{-3}$ the coronal Field's length. Therefore, we can expect that turbulent thermal conduction would be of significant importance in setting the scales over which the instability grows.





\subsection{Elongated condensations and the analogy with cloud crushing}



One of the striking features of the thermal instabilities found in these simulations is the coherency they can achieve in the $x-$ direction, as can be seen in Figure \ref{fig:volume3D}d. We have seen in Section \ref{turbTC} that turbulence can work to make the instability grow over particular scales. However, this is insufficient to explain the coherency observed in the simulations. But missing from the previous arguments is the importance of the shear flow in the $x$-direction. As a condensation forms through the thermal instability, turbulence is working to shred up the condensation (working in a way similar to thermal conduction as discussed in Section \ref{turbTC}). However, as we have a shear flow the cool material shredded from the condensation is pulled out in the $x-$ direction allowing it to act as a seed for further condensation. This drives the elongation in the $x-$ direction, i.e., in the direction of the shear flow. Because of the imposed periodic boundaries, these structures are naturally able to reach the scale of the box, but the fundamental physics that allows this to happen is not dependent on the boundary.

A similar phenomenon has been observed in modelling of other astrophysical phenomena, a key example being the process known as cloud crushing \citep{Li2019}. Cool outflowing material is commonly observed around galaxies, which is subject to strong shearing motions and hence turbulent mixing. Radiative losses associated with the mixed material can lead to cooling. The shear flow naturally takes these cooling seeds away from the initial cloud where further turbulence can result in more mixing. As a result the total amount of cool material can exceed the initial mass, as seen in 3D cloud crushing simulations \citep{Gronke2018}.



\section{Conclusions}

In this paper, the formation of cool, dense, prominence-like material has been shown to occur due to mixing-induced thermal instabilities in 3D Kelvin-Helmholtz instabilities in optically-thin radiative MHD. The thermal instabilities are highly dynamic and strongly affected by the shear, producing complex structures aligned perpendicular to the magnetic field. Approximately 20\% of all losses within the simulation are attributed to thermal instabilities. As such, thermal instabilities are an essential component for forming cool, dense material, and can self-consistently arise in the turbulent mixing layer of shearing instabilities as a result of radiative losses.


\section*{Acknowledgements}

BS and AH are supported by STFC research grant ST/V000659/1, STY00230X/1 and UKRI Research Grant No. UKRI1165. This work used the DiRAC@Durham facility managed by the Institute for Computational Cosmology on behalf of the STFC DiRAC HPC Facility (www.dirac.ac.uk). The equipment was funded by BEIS capital funding via STFC capital grants ST/P002293/1, ST/R002371/1 and ST/S002502/1, Durham University and STFC operations grant ST/R000832/1. DiRAC is part of the National e-Infrastructure.
For the purpose of open access, the author has applied a Creative Commons Attribution (CC BY) licence to any Author Accepted Manuscript version arising from this submission. 
This research was supported by the International Space Science Institute (ISSI) in Bern, through ISSI International Team project \#457 ``The Role of Partial Ionization in the Formation, Dynamics and Stability of Solar Prominences'' and ISSI International Team project 545 ``Observe Local Think Global: What Solar Observations can teach us about Multiphase Plasmas across Astrophysical Scales''. {We would also like the thank the anonymous referee for their insightful and detailed comments.}
\section*{Data Availability}

The simulation data from this study are available from BS upon reasonable request. The (P\underline{I}P) code is available at \href{https://github.com/AstroSnow/PIP}{https://github.com/AstroSnow/PIP}.



\bibliographystyle{mnras}
\bibliography{bib} 




\appendix

\section{Mass below $T_{\rm cut}$ in 2D simulations} \label{sec:appMassBelowTcut}

\begin{figure}
    \centering
    \includegraphics[width=0.95\linewidth]{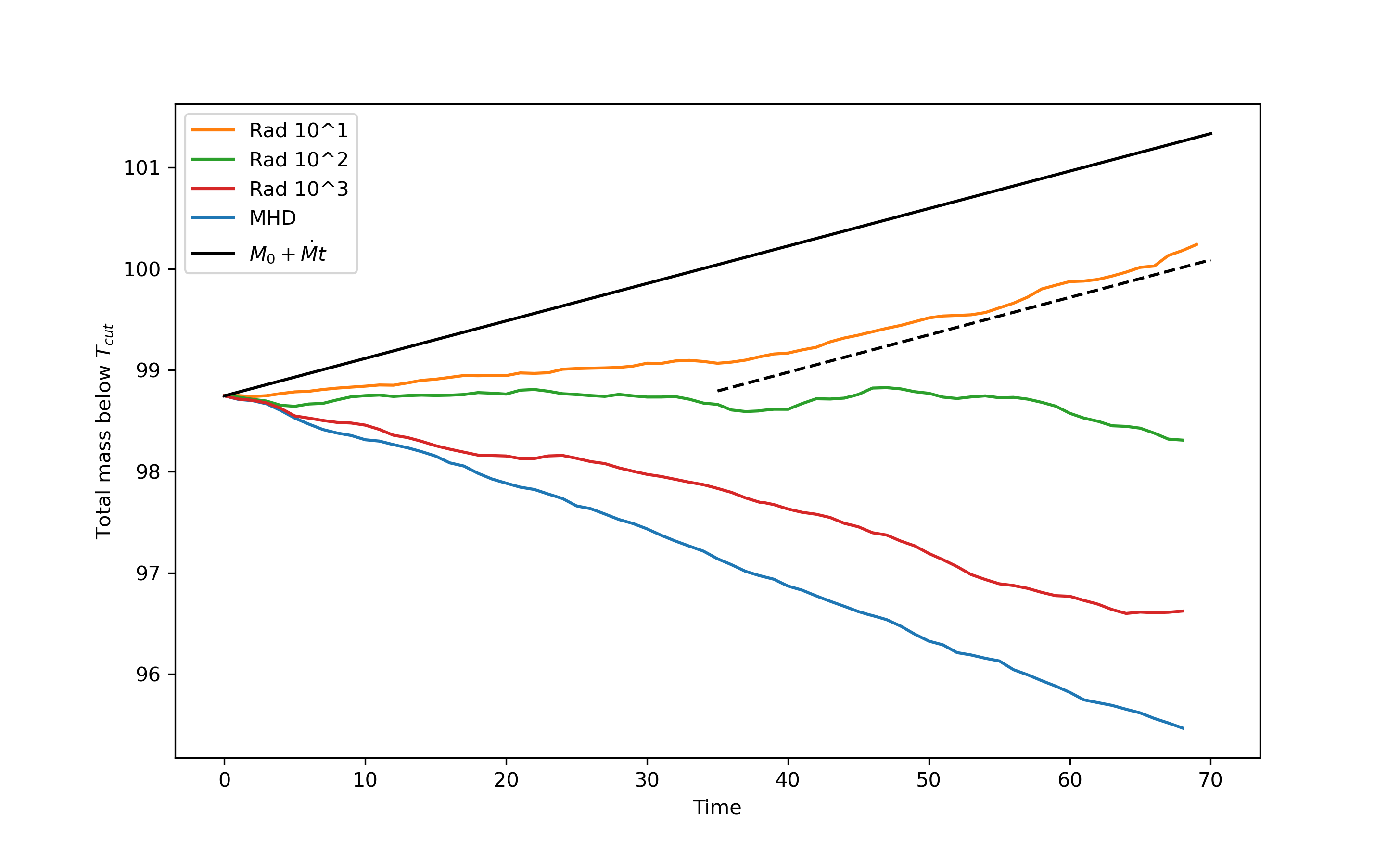}
    \caption{Mass below $T_{\rm cut}$ in the 2D simulations of \protect\cite{Hillier2023} for different radiative timescales.}
    \label{fig:massBelowTcut2D}
\end{figure}

\cite{Hillier2023} performed several 2D simulations of the radiative KHI at a condensation-corona interface, with the magnetic field in the out-of-plane direction. For different loss rates, it can be show that the trend of the total mass below $T_{\rm cut}$ varies depending on the timescale of radiative losses relative to the dynamic timescales of the system. Figure \ref{fig:massBelowTcut2D} shows that for strong losses, the mass below $T_{\rm cut}$ increases. However, as the timescale for losses increases, the mass below $T_{\rm cut}$ can stagnate. And for weak or no cooling, the total cool material mass can decrease as cool material is mixed with hot material, and the losses are not strong enough to replenish the entrained cool material. Full simulation details and results for the 2D case can be found in \cite{Hillier2023}.

\bsp	
\label{lastpage}
\end{document}